\documentclass[sigconf]{acmart}
\usepackage{graphicx,subcaption}
\usepackage{balance}
\usepackage{adjustbox}
\usepackage{amsthm}
\usepackage{enumitem}

\usepackage{tabularx}
\usepackage{multirow} 
\newtheorem{property}{PROPERTY}
\newtheorem{lem}{Lemma}
\newtheorem{ex}{Example}
\newtheorem{pf}{Proof}
\usepackage[linesnumbered,ruled,noend]{algorithm2e}
\usepackage{tikz}
\usepackage{pgfplots}
\usepackage{pgfplotstable}
\usetikzlibrary{patterns}
\usepgfplotslibrary{groupplots}
\pgfplotsset{compat=1.3,  xticklabel style = {font=\small}}

\newcommand{\Tau}{\mathrm{T}}
\newcommand{\mli}[1]{\mathit{#1}}
\setcopyright{none}
\renewcommand\footnotetextcopyrightpermission[1]{} 
\newenvironment{Function}[1][htb]
  {
  \begin{algorithm}[#1]%
  }{\end{algorithm}}

\begin{document}



\title{Automatic View Selection in Graph Databases\\ (Extended Version)}

\author{Chao Zhang}
\affiliation{%
  \institution{University of Helsinki}
  \institution{Renmin University of China}
}

\author{Jiaheng Lu}
\affiliation{%
  \institution{University of Helsinki}
}

\author{Qingsong Guo}
\affiliation{%
  \institution{University of Helsinki}
}



\author{Xinyong Zhang}
\affiliation{%
  \institution{Huawei Technologies Co., Ltd.}
}

\author{Xiaochun Han}
\affiliation{%
  \institution{Huawei Technologies Co., Ltd.}
}

\author{Minqi Zhou}
\affiliation{%
  \institution{Huawei Technologies Co., Ltd.}
}




\begin{abstract}
Recently, several works have studied the problem of view selection in graph databases. However, existing methods cannot fully exploit the graph properties of views, e.g., supergraph views and common subgraph views, which leads to a low view utility and duplicate view content. To address the problem, we propose an end-to-end graph view selection tool, G-View, which can judiciously generate a view set from a query workload by exploring the graph properties of candidate views and considering their efficacy. Specifically, given a graph query set and a space budget, G-View translates each query to a candidate view pattern and checks the query containment via a filtering-and-verification framework. G-View then selects the views using a graph gene algorithm (GGA), which relies on a three-phase framework that explores graph view transformations to reduce the view space and optimize the view benefit. Finally, G-View generates the extended graph views that persist all the edge-induced subgraphs to answer the subgraph and supergraph queries simultaneously. Extensive experiments on real-life and synthetic datasets demonstrated G-View achieved averagely 21x and 2x query performance speedup over two view-based methods while having 2x and 5x smaller space overhead, respectively. Moreover, the proposed selection algorithm, GGA, outperformed other selection methods in both effectiveness and efficiency.
\end{abstract}






\maketitle
\thispagestyle{plain}
\pagestyle{plain}



\section{Introduction}
Graph data is becoming increasingly ubiquitous across many application domains \cite{sahu2017ubiquity}, such as social networks, real-time road networks, and on-line recommendations. This trend propelled the recent proliferation of graph databases, e.g., Neo4j \cite{Neo4j} and JanusGraph \cite{JanusGraph}. One of the salient features of graph databases is the declarative graph query language \cite{fletcher2017declarative}, which enables users to succinctly query their property graphs with a wealth of distinctive features such as graph traversal and declarative pattern matching.

%
%

Materializing view is a widely used method in DBMS, which stores and reuses the query results to accelerate the similar incoming queries. When it comes to large-scale graphs, answering graph queries using materialized views can significantly save the expensive graph computation \cite{fan2014answering}. Particularly in the relational-based graph databases where the graph model is implemented upon a relational store, graph views can be utilized to speed up the queries in a native graph engine, thereby avoiding the costly relational joins for performing complex graph queries \cite{tian2019synergistic}. View selection is a well-studied topic in relational \cite{gupta2005selection,chirkova2003materializing,chaves2009towards,agrawal2000automated, yuan2020automatic}, XML \cite{katsifodimos2012materialized, mandhani2005query,  tang2009materialized}, and semantic databases \cite{goasdoue2011view, castillo2010selecting}. Various methods are proposed to select the materialized views for different target queries, e.g., SQL and XQuery \cite{xquery}. However, they are not suitable for graph view selection because they do not consider the structural properties of graph queries, e.g., subgraph patterns. It is also surprising that the amount of database research literature in graph view selection is so scarce despite graph databases have become prevalent in graph data management. Particularly, Kaskade \cite{da2019kaskade} inputs the view templates and then generates views as Cypher \cite{Cypher} queries. It modeled the view selection problem as an 0-1 Knapsack problem, and used a branch-and-bound solver to select the graph views. However, there are two major limitations to existing methods.


The first limitation is that existing methods only select views with the subgraph patterns to answer the queries while they do not consider using a view with a supergraph pattern to answer the contained queries. This leads to a low utility of the materialized views. For instance, given two view patterns and three pattern queries in Figure 1, existing methods can answer the pattern query $Q_5$ by combining the materialized results of $V_P(Q_1)$ and $V_P(Q_2)$. However, they fail to answer queries $Q_3$ and $Q_4$ despite $V_P(Q_2)$ being a supergraph pattern of them. To address this limitation, we propose an \textit{extended graph view}, which is created via an edge-induced method, being capable of answering the subgraph and supergraph queries simultaneously. Recall the example in Figure 1, with the view content $V_G(Q_1)$ and $V_G(Q_2)$, two extended graph views $V(Q_1)$ and $V(Q_2)$ can be validated to answer all the three queries $\{Q_3, Q_4, Q_5\}$. Such validation is achieved by a filtering-and-verification framework that checks the query containment by views. Furthermore, we propose a two-level search algorithm to find a minimal view set that can answer a pattern query $Q_G$ considering both subgraph and supergraph views.

\begin{figure*}[!t]
	\centering
	\includegraphics[width=1.0\linewidth]{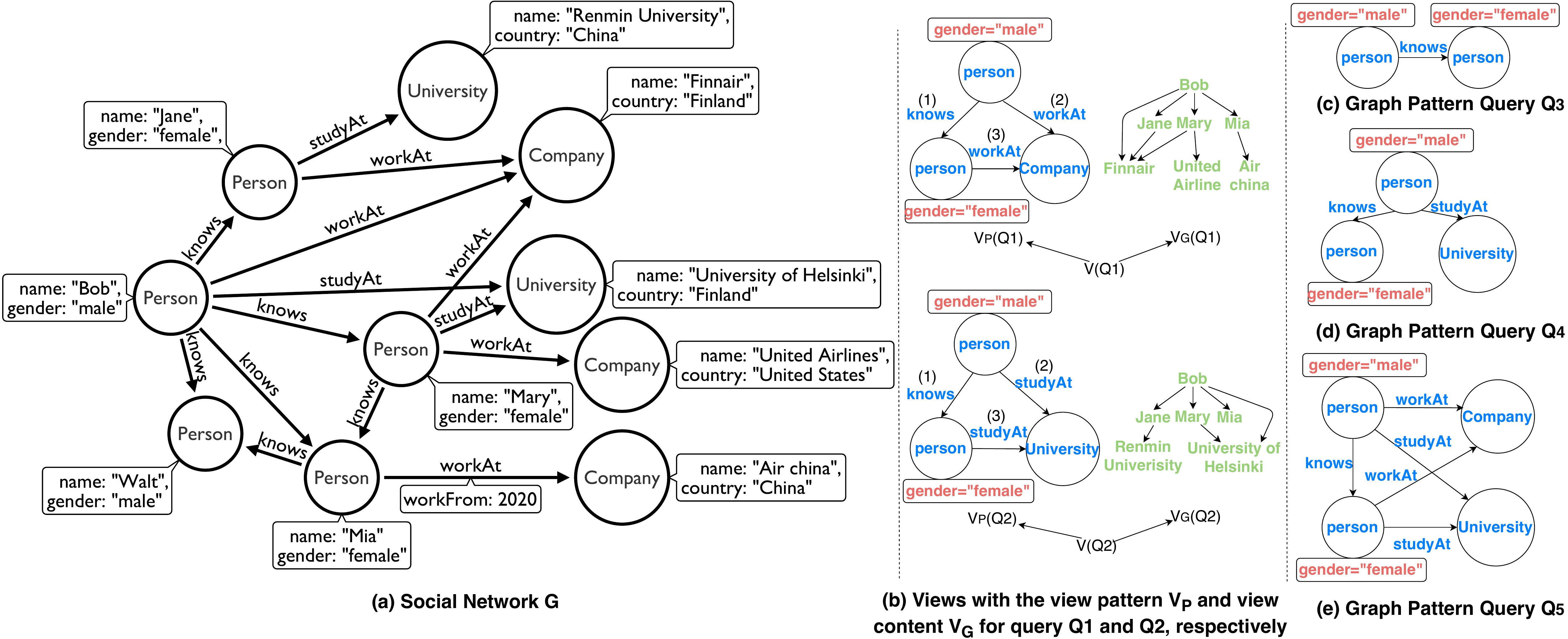}
	\caption{Labeled property graph, extended graph views, and graph pattern queries}
	\label{fig:motivatingExample}
\end{figure*}

The second limitation is that existing methods cannot effectively explore the possible candidate view combinations to reduce the view space and improve the view benefit. For instance, Kaskade \cite{da2019kaskade} can select a single view that rewrites a given query with the highest benefit, but do not consider selecting a view set $\mathcal{V}$ to rewrite a query. Such a view set $\mathcal{V}$ could be reused to answer other contained queries, thereby saving the view space. Unfortunately, generating an optimal view set $\mathcal{V}'$ for a query workload $Q$ is rather challenging due to the exponential search space. In addition, exploring the graph properties among views, e.g., finding the maximum common subgraphs \cite{minot2015comparison} to generate a smaller view set, entails an NP-hard problem of subgraph isomorphism \cite{lee2012depth}. To mitigate this problem, we propose a graph gene algorithm (GGA), which relies on a three-phase framework that heuristically explores graph view transformations to reduce the view space and optimize the view benefit. We have shown that GGA has a property of \textit{transformation completeness}, which guarantees that the query workload can be fully covered by any state of the candidate views. 

In this paper, we propose an end-to-end graph view selection tool, G-View, to judiciously generate a view set in graph databases by exploring the graph properties of candidate views and considering their efficacy. In a nutshell, given a graph query set and a space budget, G-View constructs the candidate view patterns and generate the most beneficial views to accelerate the queries. To summarize, we have made the following contributions:



\begin{enumerate}[label={(\arabic*)},leftmargin=*]
\setlength\itemsep{0em}
\item We propose an end-to-end graph view selection tool, G-View, to automatically select graph views for accelerating the graph query processing in graph databases.

\item We propose an \textit{extended graph view}, which is created by an edge-induced method that translates a graph query to a query pattern and persists all its edge-induced subgraphs to answer both subgraph and supergraph queries.

\item We propose a filtering-and-verification framework that enables the verification of the query containment by views. 


\item We propose a view selection algorithm, named GGA, to select the views into the memory under a space budget, which explores various options of graph view transformations to find an optimal view set.

\item We conducted extensive experiments on diverse query workloads and datasets. Experimental results showed that G-View can significantly accelerate the queries and reduce the overhead for other view-based methods and GGA outperformed other selection methods.
\end{enumerate}

\section{PRELIMINARIES}
This section presents the definitions of terminologies and the view selection problem. Particularly, Section 2.1 defines the property graph, pattern query, edge-induced subgraph, and extended graph view; Section 2.2 defines the view overhead and benefit; and Section 2.3 defines the view selection problem.

\subsection{Graph, Queries and Views}
\noindent \textbf{Labeled property graph.} A labeled property graph is a multi-relational, attributed, digraph $G= (V_G, E_G, L, P)$, where (1) $V_G$ is a set of vertices; (2) $E_G \subseteq V_G \times V_G$, in which $(v, v')$ denotes an edge from vertex $v$ to $v'$; (3) L is a label function such that for each vertex $v \in V_G$ (resp. edge $e \in E_G$),  L($v$) (resp. L($e$)) is a label from an alphabet $\Sigma$; (4) P is a function such that for each node $v \in V_G$ (resp. edge $e \in E_G$), P($v$) (resp. P($e$)) is a set of key/value pairs called properties. Intuitively, L indicates the type of a vertex, e.g., person, organization; P specifies the properties of a vertex, e.g., name, age, gender, or the properties of an edge, e.g., a timestamp.

\noindent \textbf{Graph pattern query.} A graph pattern query is a digraph $Q_G = (V_p, E_p, L, f)$ over a labeled property graph $G$, where (1) $V_p$ is a set of query nodes and $E_p$ is a set of query edges, respectively; (3) L is a label function such that for each vertex $v \in V_p$ (resp. edge $e \in E_p$),  L($v$) (resp. L($e$)) is a label from an alphabet $\Sigma$; (4) f is a function such that for each vertex $v \in V_p$ (resp. edge $e \in E_p$),  f($v$) (resp. f($e$)) is a Boolean predicate. (See Figure 1c for an example of the graph pattern query, the labels and predicates are marked in blue and pink color, respectively).

\noindent \textbf{Edge-induced subgraph.} An edge-induced subgraph is a graph $S= (V_S, E_S, L, P)$ that contains a subset $E_S$ of the edges of a graph $G$ together with any vertices $V_S$ that are their endpoints. Two edge-induced subgraphs of the social network $G$ are depicted in Figure 1b with green color; we use the names to represent the nodes and omit the edge labels for simplicity.


\noindent \textbf{Extended graph view.} An extended graph view is a view $V=(V_P,V_G)$, where (1) $V_P$ is a view pattern of a graph pattern query $Q_G$ with a traversal order of the edges; (2) $V_G$ is the view content that includes all the edge-induced subgraphs $S$ in the traversal order of $V_P$. (See Figure 1b for an example of the extended graph views, the edges are annotated with the traversal orders and edge labels). Note that the view content is derived from the graph $G$ by incrementally adding the matches of the graph patterns. The selected views will be materialized in the format of GraphML \cite{Tinkerpopdoc}.  In the following sections, we interchangeably use $\mli{V}$ or $V(Q_G)$ to denote an extended graph view.





\begin{ex}
Figure \ref{fig:motivatingExample}a shows a social network $G$ from LDBC \cite{erling2015ldbc}, which consists of three labels of vertices, i.e., person, university, and company, and three labels of edges, i.e., knows, studyAt, and workAt. Each vertex or edge has empty, one, or two properties. Figure \ref{fig:motivatingExample}b illustrates two extended graph views $\{V(Q1)$, $V(Q2)\}$ with view patterns and view content. Figure \ref{fig:motivatingExample}c depicts three pattern queries $\{Q_3, Q_4, Q_5\}$. It can be seen that (1) $V_P(Q_2)$ contains $Q_3$ and $Q_4$, and (2) $Q_5$ is contained by a merged pattern of $V_P(Q_1) \cup V_P(Q_2)$, and our verification method in Algorithm 1 ensures all the three pattern queries can be answered by $V(Q1)$ and $V(Q2)$ without accessing the graph $G$.


\end{ex}



\subsection{View Overhead and Benefit}
\noindent \textbf{Overhead of a materialized view.} Materializing views will trade affordable space and computation overhead for the performance gains of queries. Hence, the overhead of a materialized view includes the space overhead $s(v)$ and the computation overhead $o(v)$ of generating the view. In particular, we define $s(v)$ as the byte size occupied by a view $v$, and $o(v)$ as the CPU time and I/O cost for constructing a view $v$. 


\noindent \textbf{Benefit of a view.} Using views to answer a query can significantly accelerate the expensive queries. This is most notable when the graphs are stored in the underlying RDBMSs, where graphs are computed by joining multiple tables. Therefore, views can greatly benefit such queries. Let \textit{V} be a candidate view for the given workload $Q$, we define the view benefit $b$ as follow:
\begin{definition}
\label{def1}
\textbf{(Benefit of a view)}: Given a query workload $Q$, the benefit \textit{b} of a view \textit{V} is defined as the total cost savings by processing the queries using the view \textit{V} compared to using the graph $G$:

\small
\begin{equation}
\label{equation1}
{b(V, Q)}= \sum_{q \in Q}(w_i \times (cost(q|G)-cost(q|V)))
\end{equation}

\normalsize
\noindent where $w_i$ is the weight or frequency of query $q_i$ in Q; \textit{cost($q|G$)} and \textit{cost($q|V$)}, denote the cost of query evaluation over the graph $G$ and view $V$, respectively. The \textit{cost($q|G$)} is calculated depending on the underlying store, e.g., graph store or relational store.
\end{definition}

\noindent \textbf{Benefit of multiple views.} Using multiple views to answer a query is also possible when the view set can constitute a supergraph pattern of the query and the combined view content contains all the query results. Therefore, we define the benefit of multiple views as follows:


\begin{definition}
\label{def4}
\textbf{(Benefit of multiple views)}: Given a query $q$, the benefit \textit{b} of a multi-view set $\mathcal{V} = \{ v_1,v_2,\dots,v_m \}$ is defined as the cost savings by processing the query using the view set $\mathcal{V}$ compared to using the graph $G$: 

\small
\begin{equation}
\label{equation2}
{b(\mathcal{V}, q)}= cost(q|G) - (\sum_{V \in \mathcal{V}}cost(q|V) + cost(V_1 \bowtie \dots  \bowtie V_n)) 
\end{equation}
\normalsize

\noindent where $cost(q|G)$ denotes the query cost over the graph $G$; cost $\sum_{V \in \mathcal{V}}cost(q|V)$ is the sum of their partial evaluation cost and $cost(V_1 \bowtie \dots  \bowtie V_n)$ is the cost of combining the partial results. 
\end{definition}


\subsection{View Selection Problem}
Given a query workload $Q$ and a space budget $S$, we aim to automatically select an optimal view set $\mathcal{V}$ to materialize under the budget $S$. Therefore, the view selection problem can be modeled as a Knapsack problem of maximizing the view benefit under the space budget. We adopted a setting where the materialization cost is approximated by the view size. Such a setting assumes a cost model of view materialization that is proportional to the view size.

\begin{definition}
\textbf{(View selection problem)}: Given a workload $Q$ and a space budget $S$, the objective is to select a set of views $\mathcal{V}_s$ derived from a candidate view set $\mathcal{V}$ that fully covers the query results of $Q$, with the goal of maximizing the total benefit of $b(\mathcal{V}_s, Q)$, under the constraint that the total space occupied by $\mathcal{V}_s$ is no greater than $S$.
\label{def:View-Selection-Problem}
\end{definition}


The view selection problem is NP-hard \cite{chirkova2012materialized} for a static single-view case in which $\mathcal{V}_s$ is a subset of $\mathcal{V}$, and each view $V \in \mathcal{V}_s$ is independent so that each query $q \in Q$ is answered by a single view $V \in \mathcal{V}_s$. For such a case, there is a straightforward reduction from the Knapsack problem: find a set of k items with the space occupancy $s_1, \dots, s_k$ and the benefits $b_1, \dots, b_k$ so as to maximize the sum of the benefits of the selected items that satisfy the space budget $S$. Moreover, there could be the dynamic cases in which the views in $\mathcal{V}_s$ can be changed, e.g., by merging, breaking, and removing views. The problem in such cases becomes harder since the space of the candidate view set is extremely huge and it is unfeasible to explore all possible combinations. In addition, for the dynamic case, views are not independent as a query can be answered by multiple views, resulting in a more complicated problem than the static case using the single-view evaluation. In this work, we propose a graph gene algorithm to address the view selection problem in the dynamic multi-view setting.


\section{System Overview}
\setlength{\belowcaptionskip}{-6pt}
\begin{figure}
	\centering
	\includegraphics[width=0.85\linewidth]{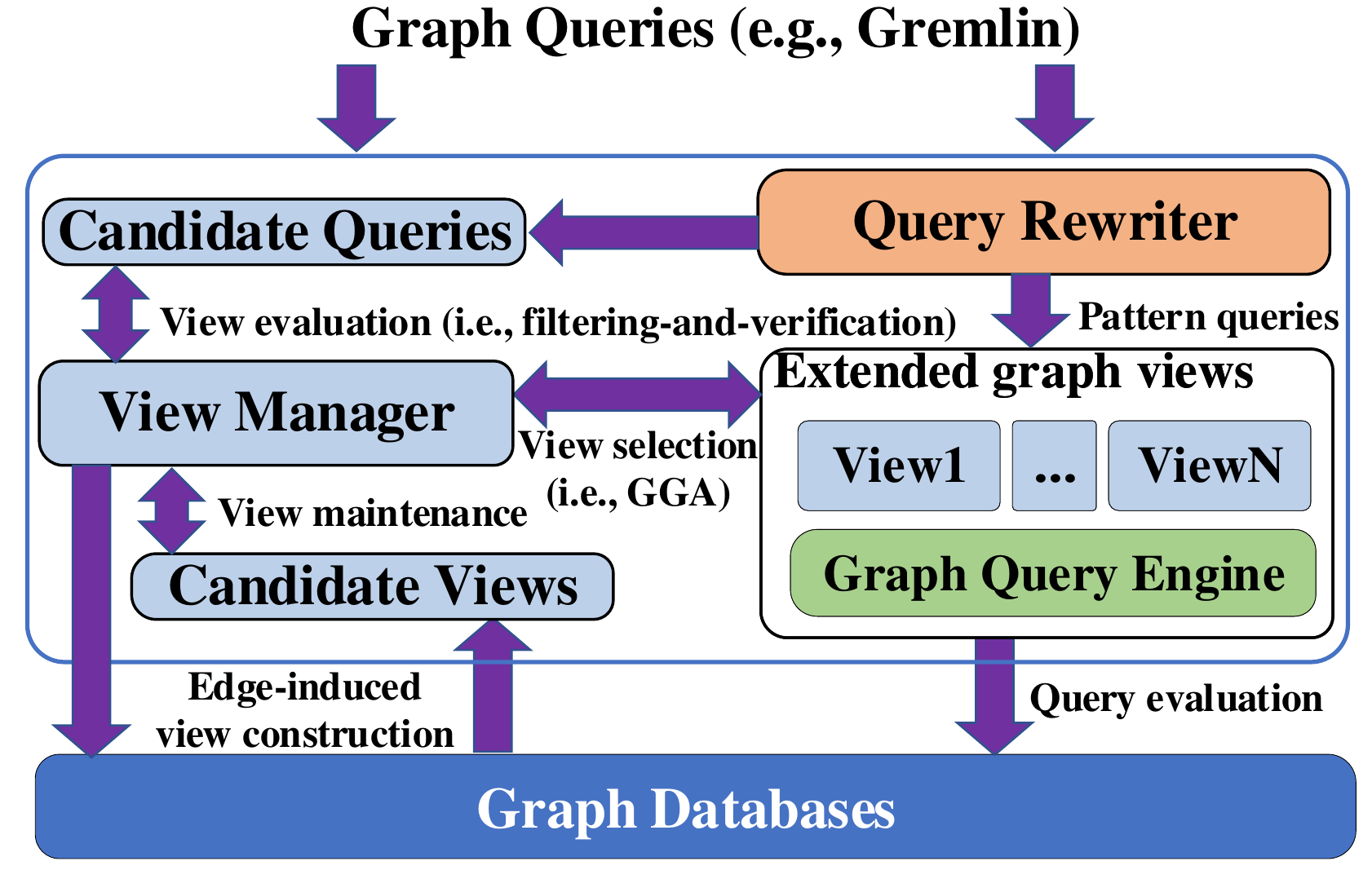}
	\caption{An overview of G-View's architecture.}
	\label{fig:overview}
\end{figure} 
In this section, we introduce the system architecture of G-View and its key components. We particularly present the implementation details for Gremlin \cite{rodriguez2015Gremlin}, which is a widely used graph query language. Figure \ref{fig:overview} shows the overall architecture of G-View. The cornerstone of the system is a view-based middle layer that is built on top of the graph databases, which accepts a set of Gremlin queries and constructs one or more extended graph views that can be utilized to speed up the queries.

The core component of G-View is the \textbf{view manager}, which has three main tasks. The first task is to verify the query containment and evaluate the view benefit for the candidate views. The methods in detail are presented in Section \ref{sec:filtering_verification}. Note that to evaluate the view benefit, the queries will also be sent to the underlying graph database for evaluation. If the graph database is implemented by an RDBMS, the Gremlin queries will eventually be translated to SQL queries based on the Gremlin2SQL technique \cite{SQLG}. The second task is to select the views based on the graph gene algorithm proposed in Section \ref{sec:selectionAlgorithms}. The third task is to generate the extended graph views from the underlying graph store via the edge-induced method introduced in Section 4.2.2. 


The \textbf{query rewriter} component translates the Gremlin queries to pattern queries based on the method in Section 4.2.1. Translating a simple pattern matching of Gremlin to a pattern query is straightforward as shown in Example 2. For future work, we plan to support other kinds of operations, such as map, filter, side effect, and branch. Conceptually, the rewriter can map other graph traversals such as linear, nested, and path traversals to pattern queries as the pattern matching query is a general traversal for Gremlin \cite{rodriguez2015Gremlin}. Concerning more expressive queries such as regular path queries (RPQs), Gremlin now supports limited RPQs \cite{angles2017foundations} by the use of \textit{repeat} step, thus many simple RPQs could be expressed as bounded pattern queries \cite{fan2014answering} for containment checking and query evaluation.


Below the query rewriter is the \textbf{extended graph view} component, which generates a set of selected views for answering the queries in a native graph engine. We adopt the TinkerGraph \cite{TinkerGraphdoc} as the in-memory graph engine coupled with the TinkerPop3 framework \cite{Tinkerpopdoc}. The view data will be stored in TinkerGraph \cite{TinkerGraphdoc} using the index-free adjacency structure \cite{lissandrini2018beyond}. By mapping the query and view to a graph pattern, it leverages a filtering-and-verification framework (see Section \ref{sec:filtering_verification}) to determine whether or not the query is contained by the views.


\begin{ex}
Consider a query in a social network of LDBC \cite{erling2015ldbc}, which finds the male persons' female friends, and the companies the friends worked at, as well as the universities the friends studied at. The corresponding Gremlin query is expressed as follows:

\ \ \ \ \ \ \ \ g.V().has(‘gender',‘male').as(‘p').match(

\ \ \ \ \ \ \ \ \underline{\hspace{0.9em}}.as(‘p').out(‘knows').as(‘f').has(‘gender',‘female'),

\ \ \ \ \ \ \ \ \underline{\hspace{1em}}.as(‘f').out(‘workAt').as(‘c'),

\ \ \ \ \ \ \ \ \underline{\hspace{1em}}.as(‘f').out(‘studyAt').as(‘u'))

\ \ \ \ \ \ \ \ .select(‘p', ‘f',‘c',‘u')

\noindent where the above query is a pattern query $Q_G = (V_p, E_p, L, f)$ which defines a set of nodes $V_p$ and edges $E_p$ in the \texttt{match} step. Particularly, each \texttt{as} step refers to a query node $v$ with a unique alias that has the mapping label $L(v)$; each \texttt{has} step defines a Boolean predicate $f(v)$ with a key-value pair; each \texttt{out} step declares an outgoing labeled edge $e$; the \texttt{select} step returns all the matched vertices. 

\end{ex}

\section{Candidate View Construction and Evaluation}
In this section, we introduce how to construct the candidate view patterns and how to create the view content, as well as how to evaluate the view benefit.

\subsection{Edge-Induced View Construction}
\label{sec:view_construction}

\setlength{\belowcaptionskip}{-3pt}
\begin{figure}
	\centering
	\includegraphics[width=0.9\linewidth]{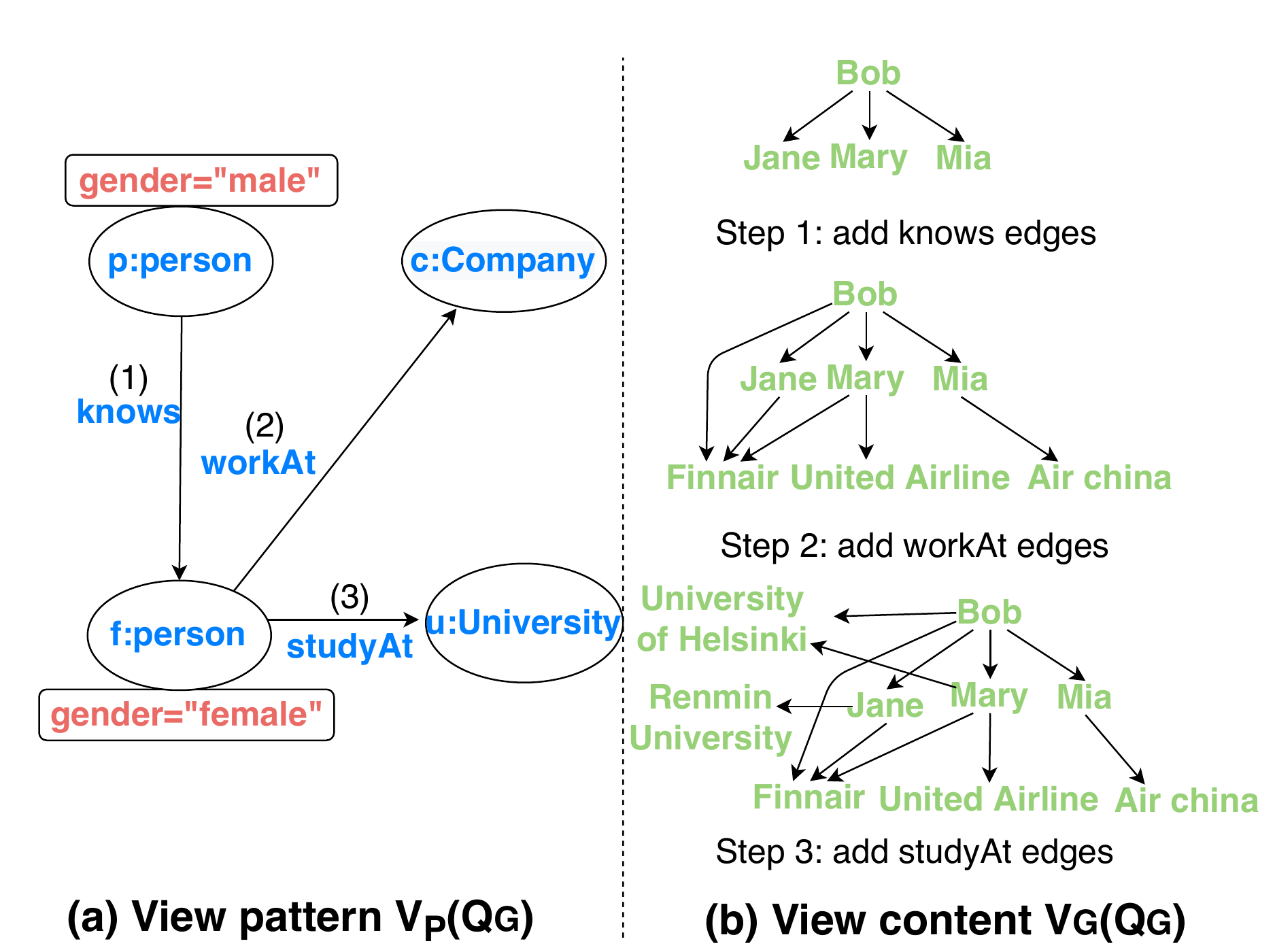}
	\caption{An example of edge-induced view construction.}
	\label{fig:induced-edges}
\end{figure} 

\subsubsection{View pattern construction.} Given a candidate query set $Q$, we translate the queries to a pattern query set, then leverage an edge-induced method to construct a candidate view for each pattern query. Particularly, for a pattern query $Q_G \in Q$, we parse it to Gremlin traversals and derive the traversal patterns $E_p$; we then add each query edge $e \in E_p$ with the predicates to its view pattern $V_P(Q_G)$ in succession. Since the query node $v$ and edge $e$ are labeled with a given alias, the procedure will also map the alias label to the label $L(v)$ and $L(e)$ in the schema graph. Consider an example in Figure \ref{fig:induced-edges}a, we derive the \textit{knows} edge as the first traversal and add it to the view pattern $V_P(Q_G)$, we then sequentially add the \textit{workAt} and \textit{studyAt} edges. Finally, we will map the aliases \{ \textit{p, f, c, u} \} to labels \{ \textit{person, person, company, university} \} that are inferred from the schema graph. We particularly construct the view pattern according to its optimized traversal order using the \textit{CountMatchAlgorithm} \cite{brocheler2011budget}.

\subsubsection{View content construction.} To construct the view content $V_G(Q_G)$, we create an edge-induced graph by the following steps: (i) we traverse each edge $e \in E_p$ in the traversal order as the view pattern $V_P(Q_G)$’s. (ii) for each visited query edge $e$, we add all the matched results of edges $E(e)$ in the property graph $G$ with their endpoints $V(e)$ to the view content $V_G(Q_G)$. (iii) the procedure terminates when all the patterns have been visited. Figure \ref{fig:induced-edges}b illustrates the procedure of constructing the view content. From step 1 to 3, the edge-induced method will append all the matched edges and vertices to the view content according to the traversal order of $[knows, workAt, studyAt]$. Regarding the detailed implementation, since the matched results of an edge depend on its previous traversals, we clone the previous traversals and cache the visited endpoints as the intermediate results and use them to compute the matches of the subsequent traversals. In such a way, the construction is much more efficient. To the end, the selected graph views are materialized in the format of GraphML \cite{Tinkerpopdoc}, which is an XML-based representation of a graph. 


\begin{property}
\label{property1}
The edge-induced view content $V_G(Q_G)$ is monotonically increasing as the view pattern $V_P(Q_G)$ grows.
\end{property}

Based on Property \ref{property1}, any graph view $V$ is not contained by another graph view $V'$ if $|E_p| > |E'_p|$. This is because we only append the edge-induce subgraphs to the view content $V_G(Q_G)$ when traversing the edge set $E_p$. The advantages of Property \ref{property1} are twofold: (1) we can decide the single view containment without matching the view patterns, and (2) it enables a query $Q_G$ can not only be contained by a set of subgraph views, but also can be answered by a supergraph view. In the next, we will present how to examine the query containment in detail.

\subsection{A Filtering-and-Verification Framework}
\label{sec:filtering_verification}
The filtering-and-verification framework consists of two stages. The first stage will check if a pattern query $Q_G$ is contained by a view pattern $V_P(Q'_G)$. Otherwise, the query will not be evaluated on view $V$. The second stage will further verify if the view content $V_G(Q_G)$ contains all the matched results of the given query. Intuitively, the first stage checks the containment between a query pattern and a view pattern, and the second stage verifies the containment between query results and the view content.

\subsubsection{The filtering stage} In this stage, we check if a pattern query $Q_G$ is a subgraph pattern of a view pattern $V_P(Q'_G)$. We first define the pattern containment as follows:

\begin{definition}
\label{def1}
\textbf{(Pattern containment)}: We say a pattern query $Q_G$ is contained by a view pattern $V_P(Q'_G)$, denoted by $ Q_G \subset V_P(Q'_G)$, if the following conditions hold:
\begin{enumerate}[leftmargin=*,label={(\roman*)}]
     \item there exists a subgraph isomorphism mapping $M$ from $Q_G$ to $V_P(Q'_G)$, such that $Q_G$ is a subgraph pattern of $V_P(Q'_G)$.

     \item for each query node $v \in Q_G$, the mapped node $v'=M(v)$ has the same label as that of $v$, and the Boolean predicates should be contained by the predicates of $v$.
     
     \item for each query edge $e \in Q_G$, the mapped edge $e'=M(e)$ has the same label that of $e$, and the Boolean predicates should be contained by the predicates of $e$.
 \end{enumerate}
\end{definition}

It is known that finding all the subgraph isomorphism mappings is NP-hard \cite{lee2012depth}, but there exist several practical algorithms to decide the answers in polynomial time. In this work, we employ the VF2 algorithm \cite{cordella2004sub} that runs in quadratic time for checking a pattern containment between two graph patterns. Note that the original VF2 algorithm does not consider the edge labels and predicates in the graph, thus we will check if the conditions (ii) and (iii) hold after a subgraph isomorphism $M$ is returned. Particurlay, the containment of Boolean predicates in conditions (ii) and (iii) means the scope of a Boolean predicate is contained by the others. For instance, a Boolean predicate \textit{(gender="male")} is contained by an empty predicate on the gender attribute, and a predicate \textit{(age<30)} is contained by a predicate \textit{(age<50)}.

\subsubsection{The verification stage} Given $Q_G \subset V_P(Q'_G)$, $Q_G$ can be answered by $V_G(Q'_G)$ if the following conditions hold:

\begin{enumerate}[label={(\roman*)},leftmargin=*]
     \item there exists a mapping $M$ from each query edge $e \in E_p $ to query edge $e' \in E'_p$.
      
     \item for the edge $e' \in E'_p$ that has no mapping from $e \in E_p$, if the vertex $v'_e \in e'$ has the mapping from $v_e \in e$, the node $v'_e=M(v)$ must have been visited in the prefix traversal patterns of $E'_p$.
\end{enumerate}
 
\begin{figure}
	\centering
	\includegraphics[width=1.0\linewidth]{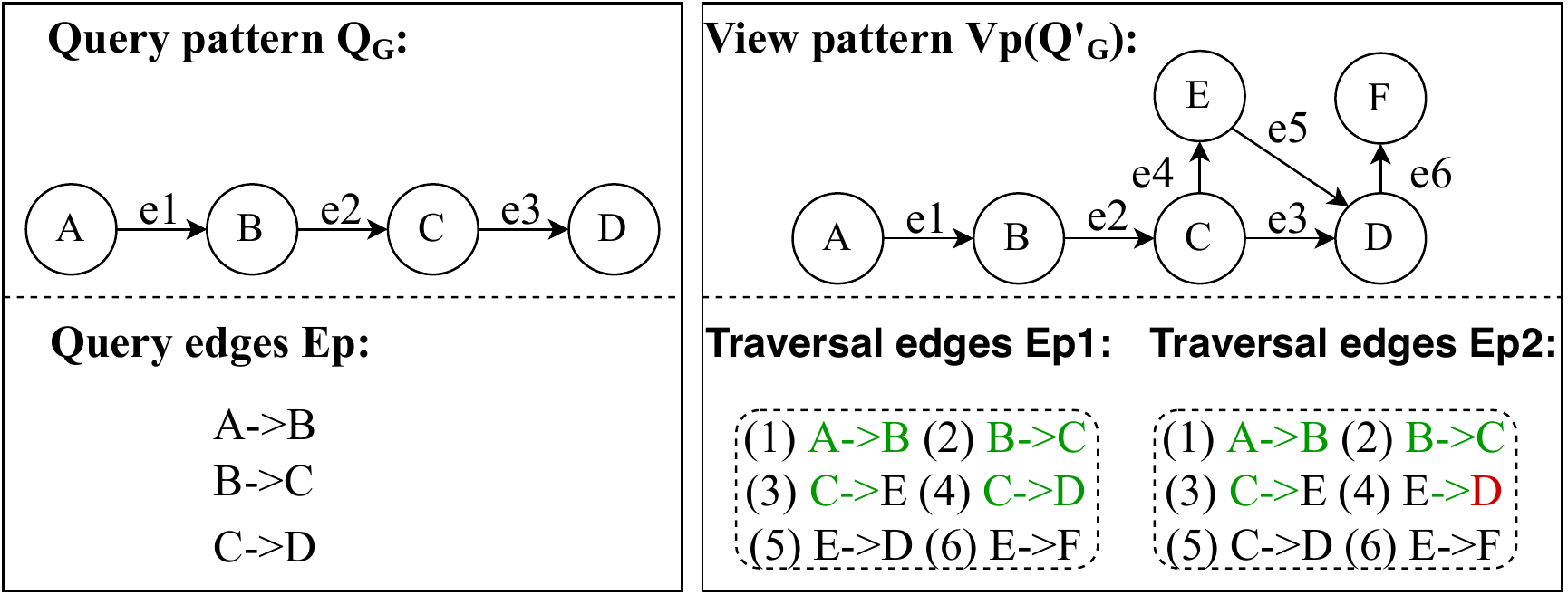}
	\caption{Containment between a query pattern and views.}
	\label{fig:filtering_verification}
\end{figure} 
 
\begin{ex}
Consider a graph pattern query $Q_G$ and a view pattern $V_P(Q'_G)$ given in Figure \ref{fig:filtering_verification}. It is clearly visible that $Q_G \subset V_P(Q'_G)$, thus the framework returns true in the filtering stage. One can verify that the $E_p$ and $E_{p1}$ satisfy the verification conditions since all the mappings of edges and vertices of $E_p$ have been visited in the prefix traversal order of $E_{p1}$. However, $E_p$ and $E_{p2}$ fail to satisfy condition (ii) in the second stage because the edge $e_5(E \rightarrow D)$ is not an edge mapping from $E_p$, and the query node $D$, which is a vertex mapping $M(v_{e3})$, has not been visited in $E_{p2}$. Therefore, it cannot guarantee that all the matched vertices of node $D$ are included.
\end{ex}
 
\begin{lem}
\label{lemma1}
The filtering-and-verification framework gives a sufficient condition to determine the query containment between the query $Q_G$ and the view \textit{V}.
\end{lem}

\begin{pf}
\textit{(Sketch)}: (1) We first prove the sufficiency of Lemma \ref{lemma1}. Given a pattern query $Q_G$ with the query edges $E_p$, and a view \textit{V} with a view pattern $V_P(Q'_G)$ and traversal edges $E'_p$, the filtering phase ensures $Q_G \subset V_P(Q'_G)$. In the second phase, if $E'_p$ has a prefix traversal pattern of $E_p$, the query $Q_G$ must be answered by the view \textit{V} because $V_G(Q'_G)$ is an edge-induced graph with all the matched results of $E'_p$ in the traversal order.

(2) We proof it is not a necessary condition by contradiction. In the verification phase, if the query nodes $V_p$ of a view pattern $V_P(Q'_G)$ follow the one-to-one relationship, the traversal order does not lead to the missing of any matched results, hence the view content $V_G(Q'_G)$ includes all the query results for $Q_G$. For instance, consider the traversal edges $E_{p2}$ in Figure \ref{fig:filtering_verification}, if both $C$ and $D$ have the one-to-one relationship to the node $E$, the view $V_G(Q'_G)$ contains all the results of $Q_G$. That concludes the proof.

\end{pf}
 
 \setlength{\textfloatsep}{3pt}%
 \begin{algorithm}[!t]
    \small
         \caption{View-based verification algorithm(VVA)}
    \KwIn{A pattern query $Q_G$, a view $V(Q'_G)$ that has a subgraph isomorphism mapping $M$ from $Q_G$.}
    \KwOut{A Boolean value deciding whether or not query $Q_G$ can be answered by $V_G(Q'_G)$.}
    \BlankLine
    $D_v \leftarrow \emptyset$, $D_e \leftarrow \emptyset$,  $l, \leftarrow E'_p.length$ \tcp*{Initialization}
    \For{$i \gets 1$ \textbf{to} $l$} { 
        \tcp{Termination condition}
          \uIf{$(D_v = V_p) \And (D_e = E_p)$}{  
            \KwRet $True$
        } 
        \tcp{Verify the condition (i)}
        \uIf {$E'_p[i]=(u'_i,v'_i) \in M(E_p)$}{
             $D_v.put(\{u'_i,v'_i\})$, $D_e.put(E'_p[i])$ 
        }
        \tcp{Verify the condition (ii)}
        \uElseIf {there is $v' \in M(V_p)$ but $E'_p[i] \notin M(E_p)$}{
            \uIf{$v' \notin D_v$}{         
            \KwRet $False$
            }
        }
    }
    \KwRet $True$\\
        \label{alg:view_verification}
\end{algorithm}
 \setlength{\textfloatsep}{5pt}%

We design a view-based verification algorithm (VVA) to decide whether or not a pattern query $Q_G$ can be answered by the view $V(Q'_G)$. As shown in Algorithm 1, it takes a pattern query $Q_G$, a view $V(Q'_G)$, and a subgraph isomorphism $M$ from the pattern query $Q_G$ to view pattern $V_P(Q'_G)$ as input, then it goes with two steps: (1) verify if condition (i) holds by checking whether or not the query edge $e' \in E'_p $ has the edge mapping in $M(E_p)$. If this is the case, add the vertex $v' \in e' $ to the dictionary $D_v$, and add the edge $e'$ to $D_e$; (2) verify if condition (ii) holds by iteratively identifying the vertex $v' \in M(V_p)$ but not in the edge mapping $M(E_p)$, and check if it has been visited. (3) the process terminates once a Boolean value has been returned. In particular, if dictionary $D_v$ has included all the vertex mappings and dictionary $D_e$ has contained all the edge mappings, then the query $Q_G$ is verified to be answered by the view content $V_G(Q'_G)$. Since checking $D_v$, $D_e$, and $M$ takes $O(1)$ time, VVA takes $O(|E'_p|)$ time to verify a query containment between a query $Q_G$ and a view $V$ where $|E'_p|$ is the size of query edges $E'_p$ of the view pattern.

\subsubsection{The applicability of extended graph views}
Concerning the applicability of the extend graph view in practice, it supports answering both the subgraph and supergraph queries with many variants based on the filtering-and-verification framework. On the one hand, since the query results are independent of the traversal orders, a query can be answered by a supergraph view only if (1) the view pattern has all the edge mappings to the query edges, and (2) the mapping vertices of the non-mapping edges in the view pattern have been visited in the traversal order (See Figure 4). This is readily done in the verification phase, which does not require the target subgraph query has the exact same traversal order as the view's. For example, the constructed view in Figure 3 with the order [(1), (2), (3)] can answer the queries with edge sets of \{(1), (2), (3)\}, \{(1), (3)\}, \{(1), (2)\} or \{(1)\} regardless of the traversal orders of the queries. On the other hand, our approach supports the multi-view rewriting that combines multiple views to answer a query, which can further significantly increase the utility of the extended views.

\subsubsection{The evaluation of view benefit}
Once the framework has verified the containment of a pattern query $Q_G$ and a view $V$, G-View then evaluates the benefit $b(Q_G,V)$. In our implementation, we use the PROFILE feature \cite{Tinkerpopdoc} of Gremlin to obtain the cost of query evaluation. Specifically, the PROFILE step returns various metrics about the given Gremlin queries including the result size, count of traversals, and total execution time in each pipeline. We perform the PROFILE step over the view \textit{V} and over the $G$, respectively, we take the total execution time as the cost and compute the benefit according to Equation \ref{equation1}.



\section{Graph Gene Algorithm}
\label{sec:selectionAlgorithms}
In this section, we propose the graph gene algorithm (GGA) for view selection. Specifically, Section 5.1 introduces the view transformations. Section 5.2 presents the evaluation method of benefit of multiple views. Section 5.3 presents the GGA algorithm.

\subsection{View Transformations}
\label{sec:4.4}

Since a view may be contained by another view, and views may have the common parts, the duplication of selected views leads to a relatively larger space occupancy and a lower coverage of the whole workload space. Based on this observation, we propose the GGA algorithm that aims at a higher usage of space and a higher benefit for the workload as a whole. The GGA algorithm is inspired by the gene algorithm (GA) \cite{beasley1993overview}, it encodes the view patterns as graph genes and solves the view selection problem as a state search process. Every state consists of a set of selected views and a total benefit. The initial state corresponds to the input candidate view set $\mathcal{V}$ with a zero benefit $b(\mathcal{V}_0)$. By merging, breaking, and removing views from the initial state, we obtain another state from view set $\mathcal{V}'$ with a new benefit $b(\mathcal{V}')$. Particularly, GGA has three atomic behaviors for view pattern transformations, namely, FISSION, FUSION, and REMOVE. GGA encodes a view pattern, a.k.a., \textit{individual}, by a set of sub-view patterns, a.k.a., \textit{graph genes}.  A new \textit{generation}, a.k.a., candidate view set, is generated by a process of probabilistic view transformations and a solution is produced based on their \textit{fitness} value, a.k.a., view benefit. In the following, we introduce the view transformations in detail.





\subsubsection{FISSION transformation.}
This transformation splits a view pattern to multiple genes. The main goal of it is to enable the identification of common parts across views heuristically as finding the common subgraphs for the graphs is an NP-hard problem. Specifically, we find the articulation points of a view pattern by using the Tarjan Algorithm \cite{tarjan1972depth}, then obtain multiple graph genes by breaking down the view pattern according to its articulation points. The articulation points are vertices whose removal increases the number of connected components of the graph, and Tarjan Algorithm \cite{tarjan1972depth} is a (Depth-First-Search) DFS-based approach that can run in \textit{O(V+E)} time to compute the articulation points in a directed graph. If the articulation point does not exist, the view pattern becomes the graph gene itself. 

\subsubsection{FUSION transformation.}
FUSION is opposite to FISSION, namely, this transformation merges or joins a view $V_i \in \mathcal{V}$ to another view $V_{j \neq i} \in \mathcal{V}$. Particularly, FUSION has two variants:

    
(1) Merge a sub-view $V_i \subset V_j$: Fusion merges the view $V_i$ to $V_{j \neq i} \in \mathcal{V}$ if $V_i$ is contained by $V_{j \neq i}$. It requires (1) $V_i$ is a subgraph of $V_{j \neq i}$; (2) $V_i$ has a prefix traversal pattern of $V_{j \neq i}$'s.
    
(2) Merge-join the genes $g_i \subset g_j$: Fusion merges the genes $g_i \in V_i$ if $g_j \in V_{j \neq i}$ contains $g_i$; the remaining genes $g_{k \neq i} \in V_i$ are joined to $V_{j \neq i}$ if they are not contained by other views $V_{k \neq i,j}$.


The first case can be decided via the filtering-and-verification framework, and contained views can be merged directly. For the second case, the algorithm enumerates all the genes $g_i \in V_i$ over the view set $\mathcal{V}$ to check the containment on other graph genes $g_j \in V_{j \neq i}$ via the filtering-and-verification framework, then merges them to the contained genes if any. The remaining genes $g_{k \neq i} \in V_i$ are assembled to the view $V_{j \neq i}$ that has contained genes by connecting the articulation points.


\subsubsection{REMOVE transformation.}
REMOVE eliminates the empty-gene candidate views after a sequence of view transformations. Such candidate views can be removed as they have been contained by other views. 


\begin{figure}[!t]
	\centering
	\includegraphics[width=0.9\linewidth]{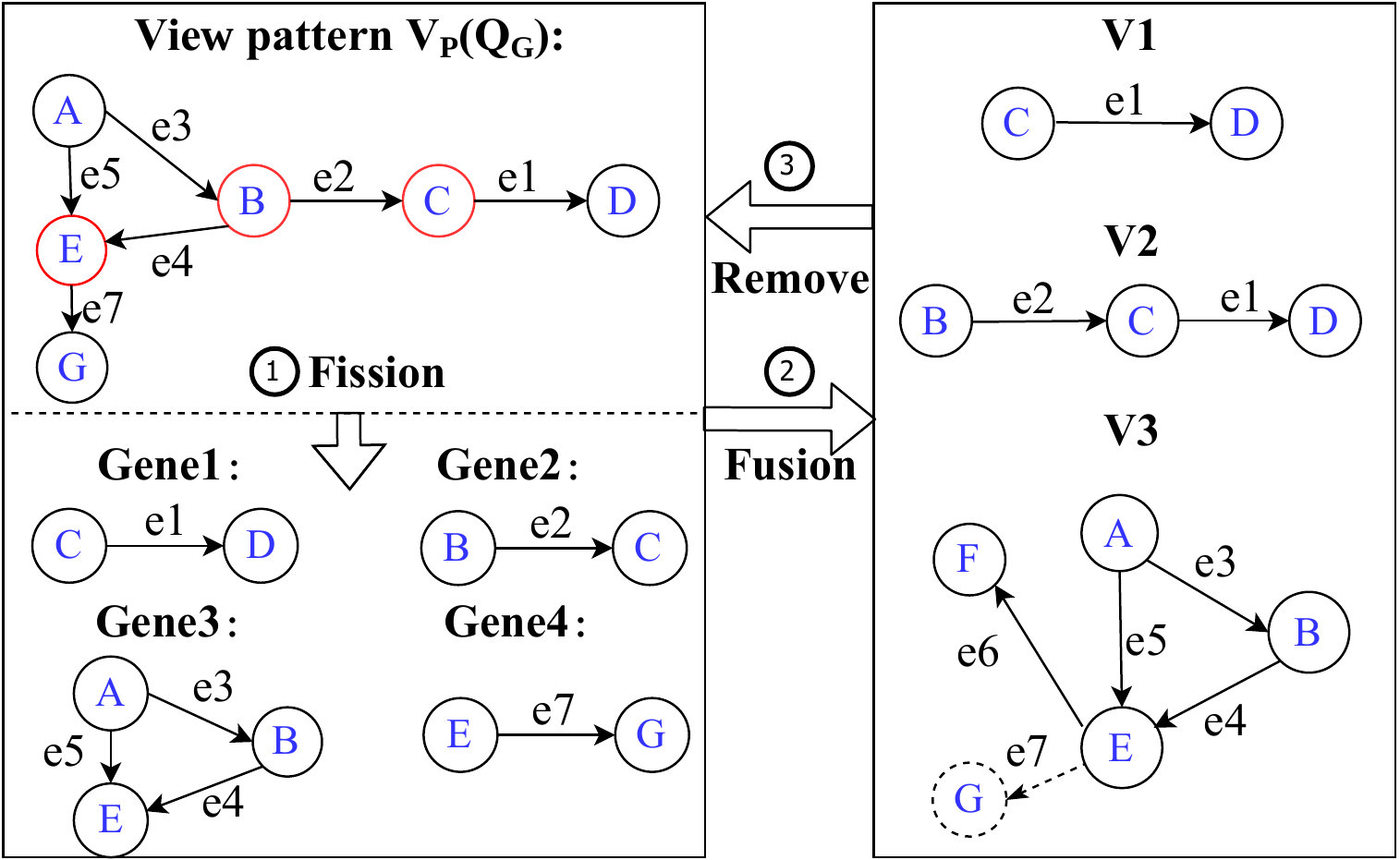}
	\caption{An illustration of view transformations.}
	\label{fig:gga}
\end{figure}

\begin{ex}
Figure \ref{fig:gga} illustrates the view transformations of the GGA algorithm. Given a view pattern $V_P(Q_G)$ and a view set $\mathcal{V}=\{V_1,V_2,V_3\}$, GGA applies a set of transformations on the view patterns. In the FISSION phase, $V_P(Q_G)$ is broken down to four genes based on the articulation points $\{B,C,E\}$. Then the genes are merged to the view set in the FUSION phase. Specifically, genes 1,2,3 are merged to $V_1,V_2,V_3$, respectively, and the remaining gene 4 is joined to $V_3$ on node E. Finally, the $V_P(Q_G)$ is removed from the candidate view set and we have reduced the common parts of three graph genes of it, i.e., genes 1, 2, and 3. 
\end{ex}






\subsection{Benefit Evaluation for Multiple Views}
\label{sec-searchandevaluate}
In this section, we introduce how to evaluate the benefit of a view set for a graph pattern query. GGA algorithm involves a case of multi-view answering, in which the graph genes have been divided and joined to different graph views (Recall the example in Figure \ref{fig:gga}). Given a pattern query $Q_G$ and a view set $\mathcal{V}$, we need to (1) find a subset of $\mathcal{V'}$ that contains $Q_G$, and (2) evaluate the total benefit $b(\mathcal{V'}, Q_G)$ and assign it to each view $V \in \mathcal{V'}$.

\subsubsection{Two-level search algorithm.} We propose a two-level search algorithm to find the minimal view set $\mathcal{V}'$ that can answer $Q_G$. Intuitively, the algorithm checks if $Q_G$ is contained by the candidate view $V \in \mathcal{V}$ in the first level, then explores the graph genes that assemble a supergraph of $Q_G$ in the second level.

\begin{algorithm}[!t]
\small
    \caption{Two-Level Minimal Search Algorithm}
    \KwIn{A pattern query $Q_G$, a candidate view set $\mathcal{V}$}
    \KwOut{A view set $\mathcal{V'}$ that minimally contains $Q_G$.}
    \BlankLine
    $\mathcal{V'} \leftarrow \emptyset$, $U_G \leftarrow \emptyset$  $M \leftarrow \emptyset$ \\
    \ForEach{view $V_i \in \mathcal{V}$}{
            \eIf{VVA($Q_G$, $V_i$)}{
                \KwRet $V_i$}
            {
                \ForEach{gene $g_{ij} \in V_i$}{
                    \uIf{VVA($Q_G$,$U_G$)}{
                        break;\\
                    }
                    \uIf{VVA($g_{Q_G}$,$g_{ij}$)}{
                        $\mathcal{V'} \gets \mathcal{V'} \cup \{V_{i}\}$ \\
                        $U_G \gets $ $U_G \cup \{g_{ij}\}$ \\
                        $M \gets M(g) \cup \{V_{i}\}$ \\
                    }
                }
                \ForEach{view $V_j \in \mathcal{V'}$}{
                    \uIf{there is no $g \in U_G$ such that $M(g)  \backslash \{V_j\} = \emptyset$ }{
                        $\mathcal{V'} \leftarrow \mathcal{V'} \backslash \{V_j\}$
                    }   
                }
            }
        }

    \KwRet $\mathcal{V'}$\\
    \label{alg:twolevel}
\end{algorithm}

Algorithm 2 depicts the two-level minimal search algorithm. Given a query $Q_G$ and a candidate view set $\mathcal{V}$ with graph genes, it returns a subset $\mathcal{V'}$ of $\mathcal{V}$ that minimally contains $Q_G$. The algorithm initializes (1) an empty set $\mathcal{V'}$ for selected views, (2) an empty set $U_G$ for merged graph genes of $\mathcal{V'}$, and (3) an index $M$ that maps each selected graph gene to a set of views (line 1). It first checks if $Q_G$ can be answered by a single view $\mathcal{V}_i$ via calling the VVA algorithm described in Algorithm 1 (lines 2-4). It will stop searching and return the view if the VVA algorithm returns true. Otherwise, it continues to find if there is any graph gene $g$ of query $Q_G$ that can be contained by the genes of $\mathcal{V}_i$. Once a qualified graph gene is found, it will add the relevant view $\mathcal{V}_i$ to the set $\mathcal{V'}$, and contained gene $g$ to the set $U_G$, respectively. Also, it will add their mapping relations to the index $M$ (lines 5-12). In lines 13-15, the algorithm removes the redundant views in $\mathcal{V'}$, such views can be eliminated as they do not any cause the missing of contained graph genes. After all the views and genes are checked, the algorithm returns $\mathcal{V'}$ (line 16). 

In the worst case that the last view is returned in the second level, the algorithm runs in O($|\mathcal{V}|(|E'_p|+|E'_p|*|g|^2)$), where $|\mathcal{V}|$ denotes the number of views in the view set $\mathcal{V}$, $|E'_p|$ is the maximum edge size of the view $ V_i \in \mathcal{V}$, and $|g|$ denotes the number of genes of the given query $Q_G$.

\begin{ex}
Consider the view pattern $V_P(Q_G)$ and view set $V=\{V_1,V_2,V_3\}$ in Figure \ref{fig:gga}. Let the genes of $V_P(Q_G)$ be $\{g_1,g_2,g_3,g_4\}$. As none of the views in the set contain $V_P(Q_G)$ in the first level, the algorithm searches for the contained genes of $V$ in the second level, where $M$ is computed to be $\{(g_1:\{V_1,V_2\}),(g_2:\{V_2\}),(g_3:\{V_3\}),(g_4:\{V_3\})\}$. As the removal of $V_1$ does not make any $M(g)$ empty, the algorithm returns $\{V_2,V_3\}$ as the minimal view set.
\end{ex}

\subsubsection{Benefit Evaluation.} After a view set $\mathcal{V'}$ that contains a query $Q_G$ is returned, we evaluate the total benefit $b(\mathcal{V'}, Q_G)$ and assign the benefit to the view set $\mathcal{V'}$ as follows: (1) compute the view cost by summing the total partial evaluation cost $\sum_{V \in \mathcal{V'}} cost(Q_G|V)$ and the cost $cost(V_1 \bowtie \dots  \bowtie V_n)$ for combining the partial results; (2) calculate the benefit by subtracting the view cost from the query cost:  $cost(Q_G|G)$; (3) finally assign the benefit to each $V \in \mathcal{V'}$ in proportion to its cost: $cost(Q_G|V)$. Note that to evaluate the partial evaluation cost, we can derive the corresponding gene $g_{Q_G} \in Q_G$, then evaluate $g_{Q_G}$ over view $V$, then we measure the combining cost by joining the views on the articulation points.



\subsection{Algorithm Description}
Integrating the methods of view transformations and benefit evaluation, we devise a view selection algorithm, called the GGA (Graph-Gene Algorithm), which is shown in Algorithm 3. Given a query workload $Q$ and a candidate view set $\mathcal{V}$, it returns a view set $\mathcal{V}_s$ that is transformed from $\mathcal{V}$. In addition, two probabilities $p_f$ and $p_f$ are provided to perform the random FISSION and FUSION transformations, respectively.

When the termination condition, e.g., a timeout threshold or an iteration number, is not satisfied, GGA repeatedly applies the FISSION, FUSION and, REMOVE transformations to derive a new state of the view selection (lines 1-8). To simplify the description, we assume the selection procedure is conducted by the helper procedure \textit{SearchAndEvaluate} (line 9), which calls the methods of benefit evaluation for multiple views in Section \ref{sec-searchandevaluate}. We use the cost optimizer \cite{TinkerGraphdoc} of Gremlin to evaluate the view benefit and In line 9. The algorithm calls another helper function \textit{DPS} to select the views based on the dynamic programming strategy.

The function of dynamic programming selection (\textit{DPS}) goes as follows: (i) initialize a benefit vector $B_V$ and a size vector $S_V$. We leverage the PROFILE \cite{Tinkerpopdoc} of Gremlin to derive the size vector $S_V$, one can also plug other size estimators, e.g., \cite{gubichev2015query}, to obtain it; (ii) fill the DP table by considering two cases for every view: (a) the view is included in the optimal subset, (b) not included in the optimal set. Therefore, the maximum value that can be obtained according to the equation: DP[i][j] = max($B_V$[i] + DP[i-1][j-$S_V$[i]],DP[i-1][j]). (iii) use a bottom-up approach to obtain the optimal selection $\mathcal{V}'$.

Note that the algorithm only jumps to a new state with a higher benefit. Otherwise, it will skip the current state and continue applying transformations to the views that are from the previously obtained state to reach a another state (lines 11-14). When the termination condition is satisfied, the algorithm returns an optimal view selection under the space budget. 





\begin{algorithm}[!t]
\small
     \caption{Graph-Gene Algorithm (GGA)}
    \KwIn{A query workload $Q$, a candidate view set $\mathcal{V}$, a space budge $S$, fission probability $p_f$, fusion probability $p_c$}
    \KwOut{A view set $\mathcal{V}_s$.}
    \BlankLine
    \While{!timeout}{
            \For{$i \gets 1$ \textbf{to} $\mathcal{V}.length$} {
             \uIf{ random(0,1) $< p_f$} {
                $\mathcal{V} = \mathcal{V}_{index \neq i} \cup FISSION(\mathcal{V}_i) $ \tcp*[f]{Fission}
             }    
             \uIf{ random(0,1) $< p_c$}{
                $\mathcal{V} = FUSION(\mathcal{V}_{index \neq i}, \mathcal{V}_i) $ \tcp*[f]{Fusion}
             }
            \uIf{ $\mathcal{V}_i.genes = empty$}{
                $\mathcal{V} = REMOVE(\mathcal{V},\mathcal{V}_i) $ \tcp*[f]{Remove}
             }
        }
        $B_v= SearchAndEvaluate(\mathcal{V}, Q)$ \tcp{Evaluation}
        $\mathcal{V}'= DPS(\mathcal{V},B_v, S)$ \tcp{View selection}
        \eIf{$B_V'>B_V$}{
            $\mathcal{V}_s = \mathcal{V}'$
        }
        {
        skip $\mathcal{V}'$\\
        }
    }
    \KwRet $\mathcal{V}_s$\\
    \label{alg:gga}
\end{algorithm}

\renewcommand{\thealgocf}{}
\begin{Function}[!t]
\small
\caption{Dynamic Programming Selection (\textit{DPS})}
    \KwIn{a candidate view set $\mathcal{V}$ with a benefit vector $B_V$, and a space budge $S$.}
    \KwOut{A subset $\mathcal{V}'$ of $\mathcal{V}$.}
    \BlankLine
    $B_V, S_V, DP[B_V][S_V] \leftarrow \emptyset$   \tcp*{Initialization}
    \ForEach{$v \in V$} 
    {
        \For{$i \gets 1$ \textbf{to} $B_V.length$} {
                \For{$j \gets 1$ \textbf{to} $S$} {
                    \eIf{$S _V[i] < j$}{         
                        $DP[i][j]=max(B_V[i]+DP[i-1][j-S_V[i]], DP[i-1][j])$ \\
                    }
                    {
                        $DP[i][j]=DP[i-1][j])$ \\
                    }
            }   
        }
        $i \leftarrow B_V.length, j \leftarrow S$ \\
        \While{$j>0$ and $i \neq 0$} {
                    \uIf{$DP[i-1][j] \neq DP[i][j]$}{         
                        $\mathcal{V}'.add(V[i])$ \\
                        $j = j - S_V[i-1]$ \\
                    }
                    $i = i-1$ \\
         }
    }
    \KwRet $\mathcal{V}'$\\
\label{function_DP}
\end{Function}

\begin{definition}
\label{def3}
\textbf{(Transformation Completeness)}: Let $\mathcal{V}$ be a set of candidate views and $\mathcal{V}^{i}$ be the $i$-th state of the candidate view set. $\mathcal{V}^{i}$ is transformation complete iff there exists a set of sequence transformation $\Tau = \{ \tau_1,\tau_2,\dots,\tau_n \}$ such that $\mathcal{V}$ and $\mathcal{V}^{i}$ cover the same workload $Q$. 
\end{definition}

\begin{lem}
\label{lemma3}
Any state of view sets in GGA algorithm is transformation complete for a candidate view set $\mathcal{V}$.
\end{lem}

\begin{pf}
\textit{(Sketch)} The transformation set T= \{FISSION, FUSION, REMOVE\} is complete for any candidate view set $\mathcal{V}$. Firstly, FISSION breaks the initial view set $\mathcal{V}$ to a fine-grained view set with graph genes. Thus, the joined view content of these graph genes can cover the view content of $\mathcal{V}$. Secondly, FUSION merges the view set $\mathcal{V'}$ with overlap genes. Hence, the union of view content of the remaining genes still covers the view content of $\mathcal{V}$. Finally, the empty-gene views are eliminated by REMOVE but they can be answered by other views. Therefore, for any state of view set, the original workload $Q$ can be covered by a new view set $\mathcal{V}$. That concludes the proof. 
\end{pf}

\begin{figure*}[!t]
    \centering
    \setlength{\overfullrule}{0pt}
    \begin{tikzpicture}

\definecolor{RYB1}{RGB}{141, 211, 199}
\definecolor{RYB2}{RGB}{255, 255, 179}
\definecolor{RYB3}{RGB}{190, 186, 218}
\definecolor{RYB4}{RGB}{251, 128, 114}
\definecolor{RYB5}{RGB}{128, 177, 211}
\definecolor{RYB6}{RGB}{253, 180, 98}
\definecolor{RYB7}{RGB}{179, 222, 105}

\pgfplotscreateplotcyclelist{colorbrewer-RYB}{
{RYB6!50!black,fill=RYB6},
{RYB2!50!black,fill=RYB2},
{RYB1!50!black,fill=RYB1},
{RYB4!50!black,fill=RYB4},
{RYB5!50!black,fill=RYB5},
{RYB6!50!black,fill=RYB6},
{RYB7!50!black,fill=RYB7},
}

\pgfplotstableread[col sep=space]{
query	SQLG	Neo4j	GRView
1	600	772	451
2	186368	921	611
3	46804	629	356
4	62585	561	326
5	10690	809	276
6	16354	13470	8054
7	400	499	150
8	102585	3804	2488
9	189	370	127
10	13426	714	406
11	16941	4066	2677
12	1131388	110219	49906
}\dataSFA
	
\pgfplotstableread[col sep=space]{
query	SQLG	Neo4j	GRView
1	220	405	210
2	12930	620	267
3	301560	1251	830
4	21012	592	512
5	16400	901	305
6	42286	1281	729
7	272259	2314	1756
8	173001	2672	771
9	285327	2863	1614
10	47578	1375	597
11	24836	1332	336
12	20122	1977	338
}\dataSFB

\pgfplotstableread[col sep=space]{
query	SQLG	Neo4j	GRView
1	10	68	6
2	106	154	31
3	105	114	38
4	84	109	38
5	496	102	15
6	696	124	91
7	895	117	5
8	2572	177	7
9	997	280	7
10	65	141	13
11	3905	571	219
12	2498	627	122
	}\dataSFC
	
	\scriptsize
	\begin{groupplot}[
		group style={group name=plots,group size=4 by 1, horizontal sep=2.5em},
		width=6.8cm,
		height=4.5cm,
		ybar=0pt,  
		xtick=data,
        tickwidth=0mm,
		enlarge x limits=0.05,
		enlarge y limits=0.05,
		xtick align=inside,
		area legend,
		legend cell align=left,
		legend columns=5,
		legend image code/.code={\draw [#1] (0cm,-0.1cm) rectangle (0.2cm,0.1cm);},
		legend style={at={(0.5,1.15)},anchor=south,fill=none},
		cycle list name=colorbrewer-RYB,
	    ymode=log
	]
	    
		\nextgroupplot[bar width=3pt,ylabel={Time(ms)},ymin=0]
		    \addplot[style = {fill=RYB6, mark=none, postaction={pattern=dots}}] table[y = SQLG] {\dataSFA};
			\addplot[style = {fill=RYB2, mark=none, postaction={pattern=north east lines}}] table[y = Neo4j] {\dataSFA};
			\addplot[style = {fill=RYB1, mark=none, postaction={pattern=horizontal lines}}] table[y = GRView] {\dataSFA};
        
		\nextgroupplot[bar width=3pt,ymin=0]
		\addplot[style = {fill=RYB6, mark=none, postaction={pattern=dots}}] table[y = SQLG] {\dataSFB};
		\addplot[style = {fill=RYB2, mark=none, postaction={pattern=north east lines}}] table[y = Neo4j] {\dataSFB};
		\addplot[style = {fill=RYB1, mark=none, postaction={pattern=horizontal lines}}] table[y = GRView] {\dataSFB};
    \legend{P-View, Subgraph, G-View}
		\nextgroupplot[bar width=3pt,ymin=0]
		    \addplot[style = {fill=RYB6, mark=none, postaction={pattern=dots}}] table[y = SQLG] {\dataSFC};
			\addplot[style = {fill=RYB2, mark=none, postaction={pattern=north east lines}}] table[y = Neo4j] {\dataSFC};
			\addplot[style = {fill=RYB1, mark=none, postaction={pattern=horizontal lines}}] table[y = GRView] {\dataSFC};
			
	\end{groupplot}
\node [text width=6cm,align=center,anchor=north] at ([yshift=-1mm]plots c1r1.south)
{\subcaption{LDBC dataset}};
\node [text width=6cm,align=center,anchor=north] at ([yshift=-1mm]plots c2r1.south)
{\subcaption{Amazon dataset}};
\node [text width=6cm,align=center,anchor=north] at ([yshift=-1mm]plots c3r1.south)
{\subcaption{DBLP dataset}};





\end{tikzpicture}
 
    \caption{Processing time on a logarithmic scale for queries, x-axis labels are query ids, i.e., Q1 to Q12.}
    \label{fig:system_performance}
\end{figure*}
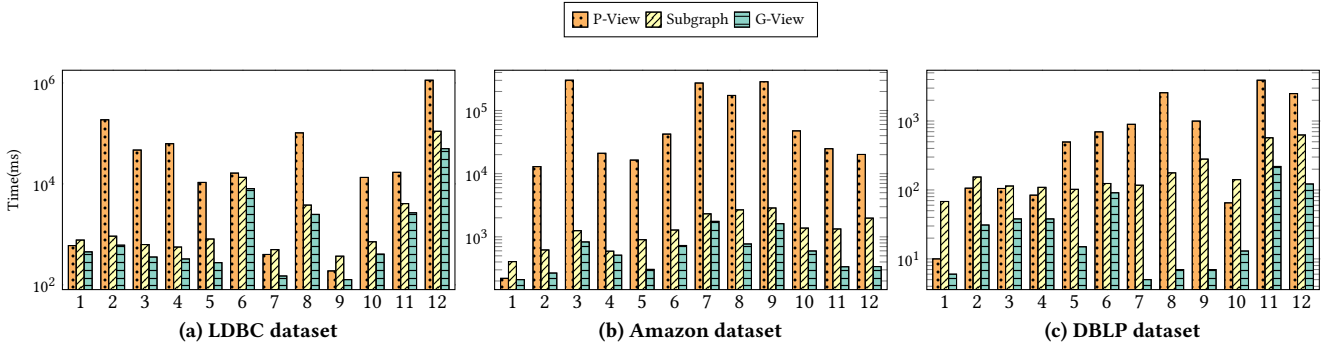

\section{Performance Evaluation}
We evaluated our methods in two aspects: (1) the query performance and materialization overhead of extended graph views and (2) the performance of the view selection algorithm, GGA.

\textbf{Compared view-based methods.} Firstly, we studied the performance using G-View against two other view-based methods. The goal is to evaluate the view benefit and overhead of the view-based methods. We constructed a graph view for each query and compared the following methods:
%

(1) \textbf{Subgraph}: We utilize the \textit{subgraph} step \cite{Tinkerpopdoc} of Gremlin to manually extract the relevant subgraphs for the queries. Although Gremlin has no explicit support for graph view, these extracted subgraphs can be treated as a form of graph view. Note that the notion of view in Subgraph is different from that of G-View as it cannot answer the queries with subgraph patterns, e.g., a view for LDBC P2 doesn't contain a view for LDBC P1 in Subgraph.     
    
(2) \textbf{Pattern views (P-View):} Fan et al. \cite{fan2014answering} proposed the graph pattern views via graph simulation \cite{fan2010graph}. The basic idea is to materialize the matches for each query edge of the graph pattern, then join the matches to answer the contained queries. Specifically, we implement it in two steps: (i) we visit the query patterns in the same traversal order as G-View's, then utilize an instance of \textit{Linkedhashmap} class in JAVA to store each query's matches, i.e., keys are query edges, value contains all the matches where each match is represented as a map from query variables to match values; (ii) we join the matches on the common keys and merge the intermediate results to answer the query at runtime.

    
    
\begin{figure}[h]
	\centering
	\includegraphics[width=1.0\linewidth]{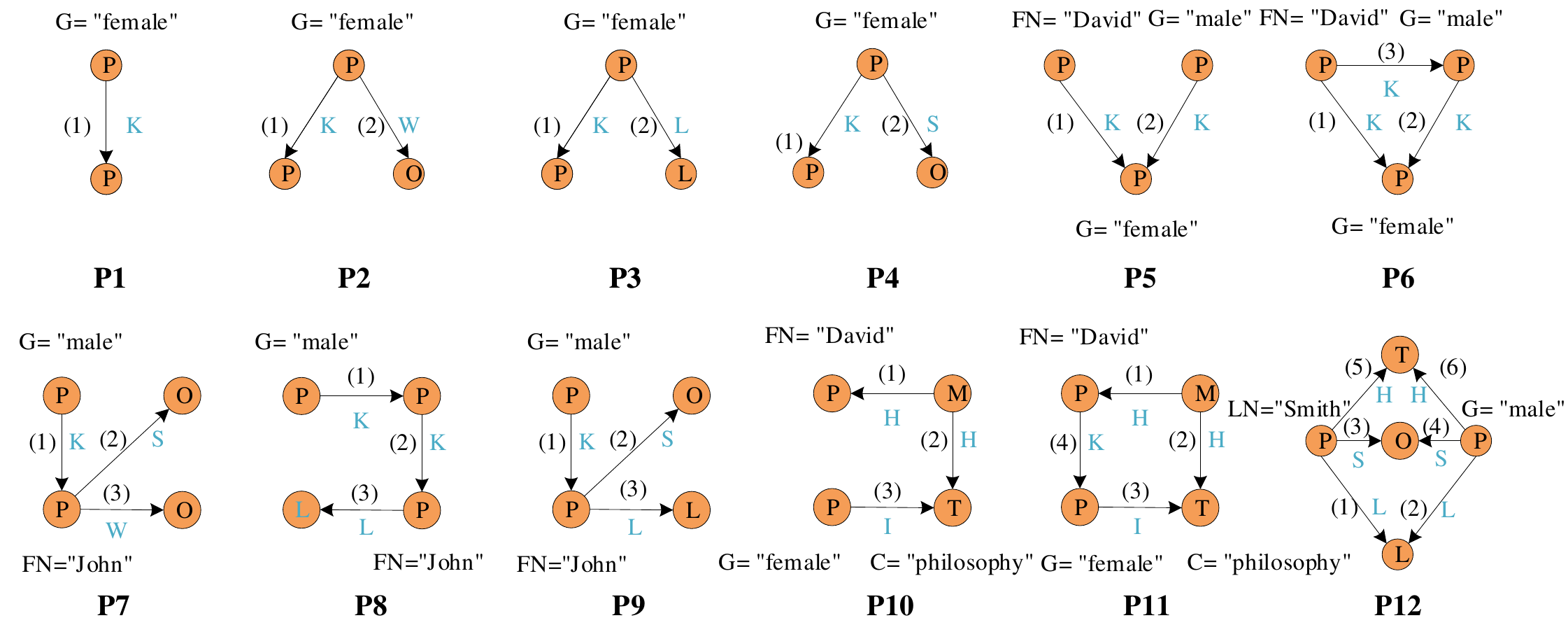}
	\caption{Graph pattern queries for LDBC dataset.}
	\label{fig:LDBC_Queries}
\end{figure}

(3) \textbf{G-View:} The third method is our method, G-View. We deployed TinkerPop v3.4.4 as the graph computing engine and used TinkerGraph to store the extended graph views. Regarding the view overhead, we also compared G-View*, which was space-optimized for G-View because of the advantage of the extended graph view supporting subgraph queries. For instance, G-View* automatically avoid constructing the views for patterns P1, P5, and P10 in the LDBC workloads (See Figure \ref{fig:LDBC_Queries}) as they have been contained by patterns P2, P6, P11, respectively.

\textbf{Compared selection algorithms.} Secondly, we measured the performance of the view selection algorithms. Specifically, all the algorithms modeled the selection problem as a Knapsack problem and they aimed at selecting the extended graph views under a space budget for a given workload. We conducted three sets of experiments to evaluate (1) the effectiveness of the algorithms in answering the query, reducing the view size, and optimizing the view benefit; (2) the efficiency of the selection algorithms; and (3) the convergence of the GGA algorithm. We compared the following selection algorithms:

(1) \textbf{Dynamic Programming Selection (DPS):} Our first baseline method is the selection method based on dynamic programming, which is described as a function in Section 5.3. 
     
(2) \textbf{Greedy-Based Selection (Greedy):} The second algorithm is a greedy-based algorithm \cite{tang2009materialized}. In particular, this method computes the view benefit in each iteration and remove a view with the maximum benefit, along with the queries it contained. The algorithm terminates until all queries are included or the total size exceeds the size constraint.

(3) \textbf{Kaskade}: The third algorithm is a branch-and-bound solver used by Kaskade \cite{da2019kaskade}. We implemented it as follows: (i) we input the view templates with no containment relationship to simulate its view enumeration; For instance, we have removed the patterns P1, P5, and P10 for the LDBC workloads; (ii) we enumerate the queries and evaluate the benefit of a view that contains the current query to simulate its single-view rewriting; (iii) we leverage the PROFILE \cite{Tinkerpopdoc} of Gremlin to derive the size vector; and finally (iv) we use a branch-and-bound solver to select the views.
    

(4) \textbf{Graph Gene Algorithm (GGA)}: The last algorithm is our view selection method, GGA, with all the details introduced in Section 5, including the methods of view transformations, multi-view evaluation, and view selection.

\setlength{\belowcaptionskip}{-1 pt}
\begin{table}[]
\begin{tabularx}{\columnwidth}{|l|X|X|X|X|X|X|X|X|}
\hline
\multirow{2}{*}{} & \multicolumn{2}{c|}{LDBC} & \multicolumn{2}{c|}{Amazon} & \multicolumn{2}{c|}{DBLP} & \multicolumn{2}{c|}{Total} \\ \cline{2-9} 
                  & $o(w)$       & $s(w)$       & $o(w)$        & $s(w)$        & $o(w)$       & $s(w)$       & $o(W)$         & $s(W)$        \\ \hline
Subgraph          &     211        &     230        &     1354         &        913      &      150       &     128        &      1715        &       1271      \\ \hline
P-View            &   190          &      128       &     1513         &        201      &      713       &     240        &      2416        &      569       \\ \hline
G-View            &     227        &      120       &       1450       &       126       &      160       &      14       &       1837       &       260      \\ \hline
\textbf{G-View*}     &      \textbf{150}       &      \textbf{103}       &    \textbf{1333}        &     \textbf{115}         &      \textbf{142}       &     \textbf{13}        &      \textbf{1625}        &      \textbf{231}       \\ \hline
\end{tabularx}
\caption{View overhead: $o(v)$ denotes the computation overhead in seconds; $s(v)$ is the space overhead in megabytes.}
\label{tab:view_overhead}
\vspace{-1mm}
\end{table}

\begin{figure}[h]
	\centering
	\includegraphics[width=1.0\linewidth]{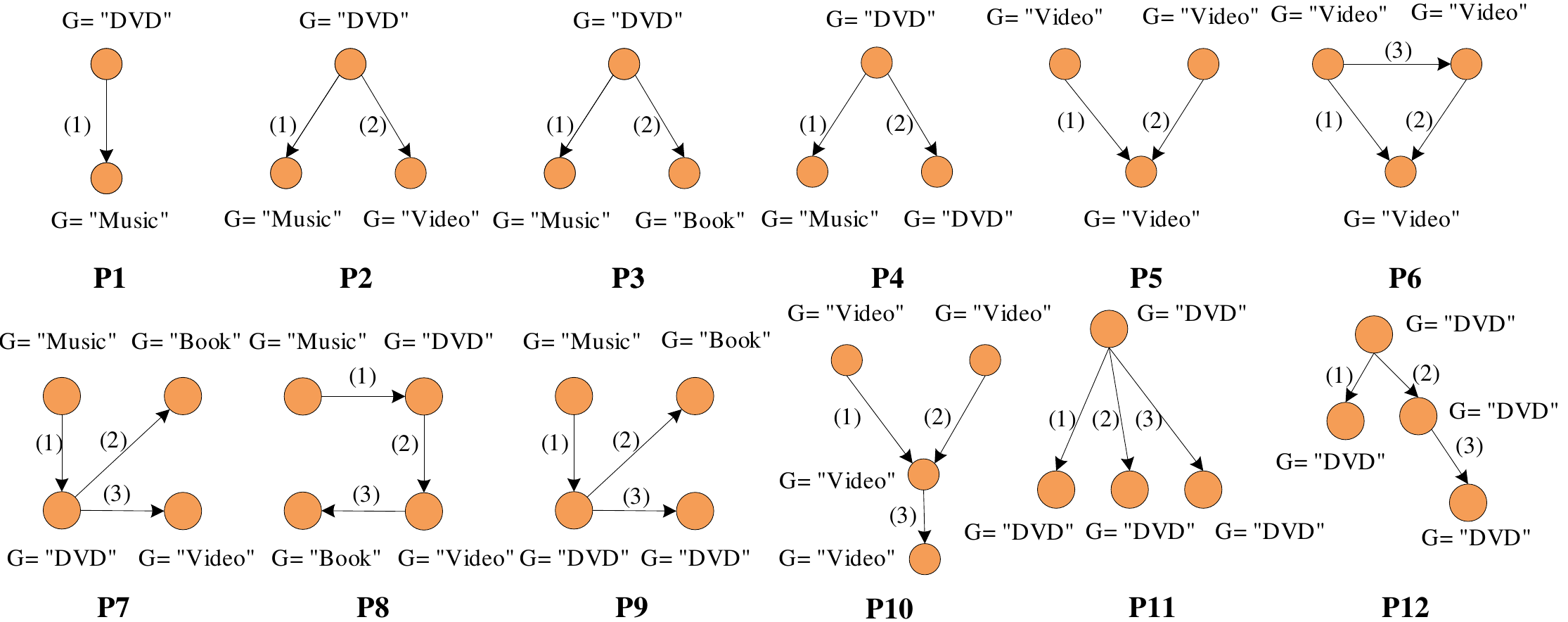}
	\caption{Graph pattern queries for Amazon dataset.}
	\label{fig:Amazon_Queries}
\end{figure}

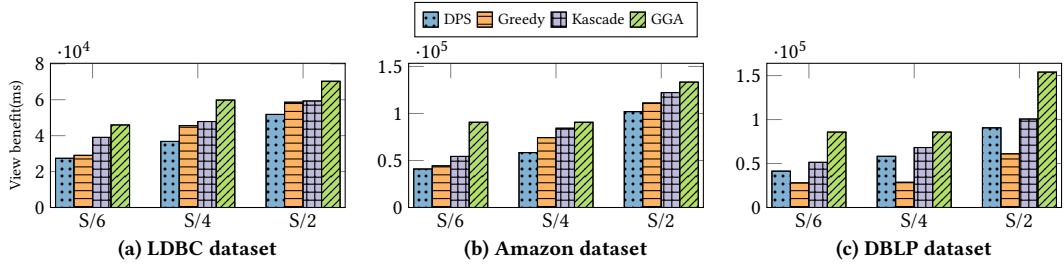
\begin{figure*}[!t]
    \centering
    \setlength{\overfullrule}{0pt}
    \begin{tikzpicture}

\definecolor{RYB1}{RGB}{141, 211, 199}
\definecolor{RYB2}{RGB}{255, 255, 179}
\definecolor{RYB3}{RGB}{190, 186, 218}
\definecolor{RYB4}{RGB}{251, 128, 114}
\definecolor{RYB5}{RGB}{128, 177, 211}
\definecolor{RYB6}{RGB}{253, 180, 98}
\definecolor{RYB7}{RGB}{179, 222, 105}

\pgfplotscreateplotcyclelist{colorbrewer-RYB}{
{RYB1!50!black,fill=RYB5},
{RYB2!50!black,fill=RYB6},
{RYB6!50!black,fill=RYB7},
{RYB4!50!black,fill=RYB4},
{RYB5!50!black,fill=RYB5},
{RYB6!50!black,fill=RYB6},
{RYB7!50!black,fill=RYB7},
}

	\pgfplotstableread[col sep=space]{
        size  DP  Greedy Kascade GGA
        1	27394	29033 39032	45957
        2	36779	45600 47874	59812
        3	51823	58611 59304	70288
	}\dataSFA
	\pgfplotstableread[col sep=space]{
        size  DP  Greedy Kascade GGA
        1	40952	44298 54314	90664
        2	58424	74312 84312	90664
        3	101951	111147 122222	133366
	}\dataSFB
	\pgfplotstableread[col sep=space]{
        size  DP  Greedy Kascade GGA
        1	41350	28022 51350	85741
        2	58208	28660 68208	85741
        3	90602	61041 100602	153967
	}\dataSFC

	\scriptsize
	\begin{groupplot}[
		group style={group name=plots,group size=3 by 1, horizontal sep=4em},
		width=5.5cm,
		height=3.5cm,
		ybar=0pt,  
		xtick=data,
        tickwidth=2mm,
		xticklabels={S/6, S/4, S/2},
		tick label style={font=\small},
		enlarge x limits=0.2,
		enlarge y limits=0,
		xtick align=inside,
		area legend,
		legend cell align=left,
		legend columns=4,
		legend image code/.code={\draw [#1] (0cm,-0.1cm) rectangle (0.2cm,0.1cm);},
		legend style={at={(0.5,1.18)},anchor=south,fill=none},
		cycle list name=colorbrewer-RYB,
	]
	    
		\nextgroupplot[bar width=7pt,ylabel={View benefit(ms)},ymin=0,ymax=80288]
	    	\addplot[style = {fill=RYB5, mark=none, postaction={pattern=dots}}] table[y = DP] {\dataSFA};
			\addplot[style = {fill=RYB6, mark=none, postaction={pattern=horizontal lines}}] table[y = Greedy] {\dataSFA};
		\addplot[style = {fill=RYB3, mark=none, postaction={pattern=grid}}] table[y = Kascade] {\dataSFA};
			\addplot[style = {fill=RYB7, mark=none, postaction={pattern=north east lines}}] table[y = GGA] {\dataSFA};

		\nextgroupplot[bar width=7pt,,ymin=0,ymax=153366]
	    	\addplot[style = {fill=RYB5, mark=none, postaction={pattern=dots}}] table[y = DP] {\dataSFB};
			\addplot[style = {fill=RYB6, mark=none, postaction={pattern=horizontal lines}}] table[y = Greedy] {\dataSFB};
			\addplot[style = {fill=RYB3, mark=none, postaction={pattern=grid}}] table[y = Kascade] {\dataSFB};
			\addplot[style = {fill=RYB7, mark=none, postaction={pattern=north east lines}}] table[y = GGA] {\dataSFB};
   
     \legend{DPS,Greedy,Kascade,GGA}
		\nextgroupplot[bar width=7pt,ymin=0,ymax=163967]
	    	\addplot[style = {fill=RYB5, mark=none, postaction={pattern=dots}}] table[y = DP] {\dataSFC};
			\addplot[style = {fill=RYB6, mark=none, postaction={pattern=horizontal lines}}] table[y = Greedy] {\dataSFC};
			\addplot[style = {fill=RYB3, mark=none, postaction={pattern=grid}}] table[y = Kascade] {\dataSFC};
			\addplot[style = {fill=RYB7, mark=none, postaction={pattern=north east lines}}] table[y = GGA] {\dataSFC};
			
	\end{groupplot}
\node [text width=6cm,align=center,anchor=north] at ([yshift=-1mm]plots c1r1.south)
{\subcaption{LDBC dataset}};
\node [text width=6cm,align=center,anchor=north] at ([yshift=-1mm]plots c2r1.south)
{\subcaption{Amazon dataset}};
\node [text width=6cm,align=center,anchor=north] at ([yshift=-1mm]plots c3r1.south)
{\subcaption{DBLP dataset}};





\end{tikzpicture}
 
    \caption{View benefit for the workloads in three datasets based on views selected by three algorithms.}
    \label{fig:view_benefit}
\end{figure*}

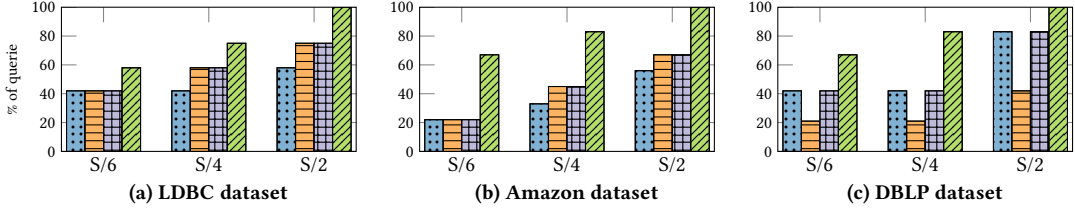
\begin{figure*}[!t]
    \centering
    \setlength{\overfullrule}{0pt}
    \begin{tikzpicture}

\definecolor{RYB1}{RGB}{141, 211, 199}
\definecolor{RYB2}{RGB}{255, 255, 179}
\definecolor{RYB3}{RGB}{190, 186, 218}
\definecolor{RYB4}{RGB}{251, 128, 114}
\definecolor{RYB5}{RGB}{128, 177, 211}
\definecolor{RYB6}{RGB}{253, 180, 98}
\definecolor{RYB7}{RGB}{179, 222, 105}

\pgfplotscreateplotcyclelist{colorbrewer-RYB}{
{RYB1!50!black,fill=RYB5},
{RYB2!50!black,fill=RYB6},
{RYB6!50!black,fill=RYB7},
{RYB4!50!black,fill=RYB4},
{RYB5!50!black,fill=RYB5},
{RYB6!50!black,fill=RYB6},
{RYB7!50!black,fill=RYB7},
}

	\pgfplotstableread[col sep=space]{
    f  BDA  GBA Kascade GGA
    1	42	42 42	58
    2	42	58 58	75
    3	58	75 75	100
	}\dataSFA
	\pgfplotstableread[col sep=space]{
    f  BDA  GBA Kascade GGA
    1	22	22 22	67
    2	33	45 45	83
    3	56	67 67	100
	}\dataSFB
	\pgfplotstableread[col sep=space]{
    f  BDA  GBA Kascade GGA
    1	42	21 	42 67
    2	42	21	42 83
    3	83	42	83 100
	}\dataSFC

	\scriptsize
	\begin{groupplot}[
		group style={group name=plots,group size=3 by 1, horizontal sep=4em},
		width=5.5cm,
		height=3.5cm,
		ybar=0pt,  
		xtick=data,
        tickwidth=2mm,
		xticklabels={S/6, S/4, S/2},
		enlarge x limits=0.2,
		enlarge y limits=0,
		xtick align=inside,
		area legend,
		legend cell align=left,
		legend columns=5,
		legend image code/.code={\draw [#1] (0cm,-0.1cm) rectangle (0.2cm,0.1cm);},
		legend style={at={(0.5,1.15)},anchor=south,fill=none},
		cycle list name=colorbrewer-RYB,
	]
	    
		\nextgroupplot[bar width=7pt,ylabel={\% of querie},ymin=0,ymax=100]
	    	\addplot[style = {fill=RYB5, mark=none, postaction={pattern=dots}}] table[y = BDA] {\dataSFA};
			\addplot[style = {fill=RYB6, mark=none, postaction={pattern=horizontal lines}}] table[y = GBA] {\dataSFA};
			\addplot[style = {fill=RYB3, mark=none, postaction={pattern=grid}}] table[y = Kascade] {\dataSFA};
			\addplot[style = {fill=RYB7, mark=none, postaction={pattern=north east lines}}] table[y = GGA] {\dataSFA};
        
		\nextgroupplot[bar width=7pt,ymin=0,ymax=100]
	    	\addplot[style = {fill=RYB5, mark=none, postaction={pattern=dots}}] table[y = BDA] {\dataSFB};
			\addplot[style = {fill=RYB6, mark=none, postaction={pattern=horizontal lines}}] table[y = GBA] {\dataSFB};
			\addplot[style = {fill=RYB3, mark=none, postaction={pattern=grid}}] table[y = Kascade] {\dataSFB};
			\addplot[style = {fill=RYB7, mark=none, postaction={pattern=north east lines}}] table[y = GGA] {\dataSFB};
    
		\nextgroupplot[bar width=7pt,,ymin=0,ymax=100]
	    	\addplot[style = {fill=RYB5, mark=none, postaction={pattern=dots}}] table[y = BDA] {\dataSFC};
			\addplot[style = {fill=RYB6, mark=none, postaction={pattern=horizontal lines}}] table[y = GBA] {\dataSFC};
			\addplot[style = {fill=RYB3, mark=none, postaction={pattern=grid}}] table[y = Kascade] {\dataSFC};
			\addplot[style = {fill=RYB7, mark=none, postaction={pattern=north east lines}}] table[y = GGA] {\dataSFC};
			
	\end{groupplot}
\node [text width=6cm,align=center,anchor=north] at ([yshift=-1mm]plots c1r1.south)
{\subcaption{LDBC dataset}};
\node [text width=6cm,align=center,anchor=north] at ([yshift=-1mm]plots c2r1.south)
{\subcaption{Amazon dataset}};
\node [text width=6cm,align=center,anchor=north] at ([yshift=-1mm]plots c3r1.south)
{\subcaption{DBLP dataset}};





\end{tikzpicture}
 
    \caption{Fraction of queries for the workloads in three datasets covered by views selected by three algorithms.}
    \label{fig:query_fraction}
\end{figure*}
\textbf{Datasets and Workloads.} We used both synthetic and real-life data to compare the performance of our approach with state-of-the-art methods. We used the data and designed the corresponding workloads as follows:

\textit{(1) Synthetic graphs.} We used a synthetic social network dataset from the LDBC benchmark \cite{erling2015ldbc}, which includes 11 entities connected by 20 relations. We generated an LDBC graph with the scale factor SF1, resulting in a graph with roughly 1M vertices and 2M edges. We designed a workload including 12 pattern queries following \cite{LdbcTechSpecification}, which are shown in Figure  \ref{fig:LDBC_Queries}. Nodes are labeled with P(Person), T(Tag), L(Location), O(Organization), and M(Message). Edges are labeled with K(Knows), Has (H), LocatedIn (L), InterestedIn (I), StudyAt (S), and WorkAt (W). Boolean predicates include Gender (G), Category (C), and Names (FN: firstName, LN: lastName). 


\textit{(2) Real-life graphs.} We used two real-life graphs: (a) Amazon dataset \cite{leskovec2007dynamics}, a product co-purchasing network with 542K nodes and 3.3M edges. Each node has attributes such as title, group and sales-rank, and an edge models a co-purchase relationship between product \textit{a} and \textit{b}. We designed 12 frequent query patterns following \cite{leskovec2006patterns}, where each of the view content contains 67K nodes and edges on average. The patterns are shown in Figure  \ref{fig:Amazon_Queries}. (b) DBLP-citation network \cite{tang2008arnetminer}, a bibliography that provides the publication information and co-authorship in the field of computer science. The dataset has 1M nodes and 2M edges, in which nodes represent papers with attributes such as title, authors, year and venue, and an edge indicate a citation from paper x to y. We also identified 12 query patterns (not shown) similar to Amazon patterns.


 

\textbf{Experimental Setup.} All the experiments were conducted on a machine with a 2-core i5 CPU (2.9 GHz) and 16GB RAM. We composed the queries using Gremlin pattern matching. We implemented all the compared methods in JAVA 1.8. We deployed SQLG v2.0.2 to stored the raw data. We constructed the views from SQLG and materialized them to GraphML \cite{Tinkerpopdoc} files.




\subsection{Evaluation of Performance and Overhead}
In this section, we evaluated the performance and overhead of three view-based methods, namely, Subgraph, P-View, and G-View. As for the view performance, we measured the running time of each query over the view in milliseconds, and we reported the running time in log scale. We ensured the consistent query results. Regarding the view overhead, we reported their computation and space overhead with respect to the query workloads.

Figure \ref{fig:system_performance} shows the evaluation results of compared methods, which clearly indicates that G-View outperformed other view-based methods in accelerating all queries. Particularly, for the LDBC dataset, it achieved 12x and 1.2x speedup for P-View and Subgraph, respectively. For the Amazon dataset, it was 69x and 2x faster. For the DBLP dataset, it achieved up to 20x and 4x speedup, respectively. Surprisingly, G-View was faster than Subgraph that is a native approach of Gremlin. We found that this was mainly because Subgraph contained many non-query results after creating the views, leading to a larger view space than that of G-View. As a result, it was slower due to a higher graph traversal cost. For example, to create a view for the Amazon P1, it will add redundant edges when applying both "G=DVD" and "G=Music" to the co-purchased edge. Instead, G-View can apply the predicates on both ends of a query edge without such side effect, resulting in a more fine-grained view. P-View has the highest computation cost due to its "relational-style" way that joins the matches of connected query edges at runtime.

Table \ref{tab:view_overhead} reports the view overheads including the computation overhead $o(v)$ and space overhead $s(v)$. Interestingly, G-View had a higher computation overhead than Subgraph's despite its space overhead was lower. We found the reason is that Subgraph is a built-in method of Gremlin, thus it can generate the view data in place without fetching the data from the graph. We believe this optimization can also be applied to G-View once it is implemented inside Gremlin. P-View had a higher space overhead than G-View's as it stored the matches for each query edge separately. Nevertheless, it had a lower space overhead than Subgraph because it can apply the predicates of a query edge simultaneously. A side observation is that P-View's computation cost was highest due to the additional transformation cost from the traversal results to pattern matches. G-View and Subgraph had no such cost as they stored the results as graphs. Last but not least, the space-optimized G-View* had the lowest computation and space overhead as it has avoided the unnecessary view construction. Another benefit of G-View is that the views are automatically generated and natively evaluated for the queries. In contrast, Subgraph has to manually construct the views and P-View evaluates the queries in a relational way.        

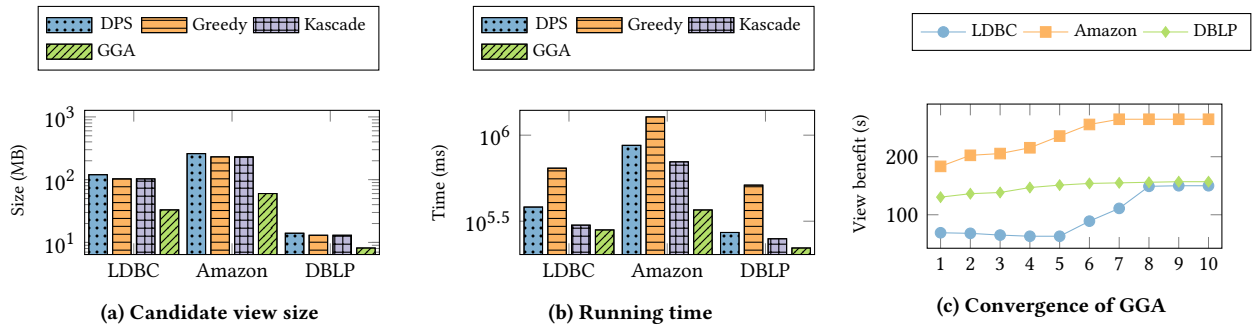
\begin{figure*}[!t]
    \centering
    \begin{subfigure}{0.31\textwidth}
    \begin{tikzpicture}

\definecolor{RYB1}{RGB}{141, 211, 199}
\definecolor{RYB2}{RGB}{255, 255, 179}
\definecolor{RYB3}{RGB}{190, 186, 218}
\definecolor{RYB4}{RGB}{251, 128, 114}
\definecolor{RYB5}{RGB}{128, 177, 211}
\definecolor{RYB6}{RGB}{253, 180, 98}
\definecolor{RYB7}{RGB}{179, 222, 105}

\pgfplotscreateplotcyclelist{colorbrewer-RYB}{
{RYB1!50!black,fill=RYB5},
{RYB2!50!black,fill=RYB6},
{RYB6!50!black,fill=RYB7},
{RYB4!50!black,fill=RYB4},
{RYB5!50!black,fill=RYB5},
{RYB6!50!black,fill=RYB6},
{RYB7!50!black,fill=RYB7},
}

\pgfplotstableread[col sep=space]{
        dataset  DPS  Greedy Kascade GGA
        1 120	103 103	33
        2 260	231 231	60.1
        3 14	13 13	8.15

}\data

\begin{axis}[
	width=5.5cm,
	height=3.5cm,
    ybar,
    label style={font=\fontsize{7}{8}\selectfont},
    bar width=7pt,
	enlarge x limits=0.25,
	enlarge y limits=0.05,
	xtick align=inside,
	area legend,
    legend style={at={(1.0,1.5)},
      anchor=east,legend columns=3, font=\fontsize{7}{8}\selectfont},
    ylabel={\ Size (MB)},
    xticklabels={LDBC, Amazon, DBLP},
    xtick=data,
    ymode=log,
    ymax=1000,
    ]
	\addplot[style = {fill=RYB5, mark=none, postaction={pattern=dots}}] table[y = DPS] {\data};
	
	\addplot[style = {fill=RYB6, mark=none, postaction={pattern=horizontal lines}}] table[y = Greedy] {\data};
	
	\addplot[style = {fill=RYB3, mark=none, postaction={pattern=grid}}] table[y = Kascade] {\data};
			
	\addplot[style = {fill=RYB7, mark=none, postaction={pattern=north east lines}}] table[y = GGA] {\data};

\legend{DPS,Greedy,Kascade,GGA}
\end{axis}






\end{tikzpicture}
 
        \caption{Candidate view size}
    \label{fig:view_size}
    \end{subfigure}
    \begin{subfigure}{0.31\textwidth}
    \begin{tikzpicture}

\definecolor{RYB1}{RGB}{141, 211, 199}
\definecolor{RYB2}{RGB}{255, 255, 179}
\definecolor{RYB3}{RGB}{190, 186, 218}
\definecolor{RYB4}{RGB}{251, 128, 114}
\definecolor{RYB5}{RGB}{128, 177, 211}
\definecolor{RYB6}{RGB}{253, 180, 98}
\definecolor{RYB7}{RGB}{179, 222, 105}

\pgfplotscreateplotcyclelist{colorbrewer-RYB}{
{RYB1!50!black,fill=RYB5},
{RYB2!50!black,fill=RYB6},
{RYB6!50!black,fill=RYB7},
{RYB4!50!black,fill=RYB4},
{RYB5!50!black,fill=RYB5},
{RYB6!50!black,fill=RYB6},
{RYB7!50!black,fill=RYB7},
}

\pgfplotstableread[col sep=space]{
        dataset  DPS  Greedy Kascade  GGA
		1 382171	643047 300000	281429
        2 872515	1277079 700000	367659
        3 271961	512424	250000 221017
}\data

\begin{axis}[
	width=5.5cm,
	height=3.5cm,
    ybar,
    label style={font=\fontsize{7}{8}\selectfont},
    bar width=7pt,
	enlarge x limits=0.25,
	enlarge y limits=0.05,
	xtick align=inside,
	area legend,
    legend style={at={(1.0,1.5)},
      anchor=east,legend columns=3, font=\fontsize{7}{8}\selectfont},
    ylabel={Time (ms)},
    xticklabels={LDBC, Amazon, DBLP},
    xtick=data,
    ymode=log,
    ]
		    \addplot[style = {fill=RYB5, mark=none, postaction={pattern=dots}}] table[y = DPS] {\data};
			\addplot[style = {fill=RYB6, mark=none, postaction={pattern=horizontal lines}}] table[y = Greedy] {\data};
			
			\addplot[style = {fill=RYB3, mark=none, postaction={pattern=grid}}] table[y = Kascade] {\data};
			\addplot[style = {fill=RYB7, mark=none, postaction={pattern=north east lines}}] table[y = GGA] {\data};

\legend{DPS,Greedy,Kascade,GGA}
\end{axis}






\end{tikzpicture}
 
    \caption{Running time}
    \label{fig:running_time}
    \end{subfigure}
    \begin{subfigure}{0.31\textwidth}
    \begin{tikzpicture}

\definecolor{RYB1}{RGB}{141, 211, 199}
\definecolor{RYB2}{RGB}{255, 255, 179}
\definecolor{RYB3}{RGB}{190, 186, 218}
\definecolor{RYB4}{RGB}{251, 128, 114}
\definecolor{RYB5}{RGB}{128, 177, 211}
\definecolor{RYB6}{RGB}{253, 180, 98}
\definecolor{RYB7}{RGB}{179, 222, 105}

\pgfplotscreateplotcyclelist{colorbrewer-RYB}{
{RYB1!50!black,fill=RYB5},
{RYB2!50!black,fill=RYB6},
{RYB6!50!black,fill=RYB7},
{RYB4!50!black,fill=RYB4},
{RYB5!50!black,fill=RYB5},
{RYB6!50!black,fill=RYB6},
{RYB7!50!black,fill=RYB7},
}

\pgfplotstableread[col sep=space]{
        dataset  LDBC  Amazon  DBLP
        1	69	183.370	130.352
        2	68	202.410	136.273
        3	65	205.420	138.370
        4	63	215.420	146.899
        5	63	235.542	151.075
        6	89	255.545	153.967
        7	111	264.545	154.967
        8	149	264.545	155.967
        9	150	264.548	156.967
        10	150	264.548	156.967
}\data

\begin{axis}[
	width=5.5cm,
	height=3.5cm,
    label style={font=\fontsize{7}{8}\selectfont},
	enlarge x limits=0.05,
	enlarge y limits=0.1,
	xtick align=inside,
    legend style={at={(1.1,1.5)},
      anchor=east,legend columns=3, font=\fontsize{7}{8}\selectfont},
    ylabel={\ View benefit (s)},
    xtick=data,
    ymax=270,
    ]
		    \addplot[RYB5,mark=*] table[y = LDBC] {\data};
			\addplot[RYB6,mark=square*] table[y = Amazon] {\data};
			\addplot[RYB7,mark=diamond*] table[y = DBLP] {\data};

\legend{LDBC,Amazon,DBLP}
\end{axis}






\end{tikzpicture}
 
    \label{fig:convergence}
    \caption{Convergence of GGA}
    \end{subfigure}
    \caption{View size, running time and converge of selection algorithms.}
    \label{fig:EX3}
\end{figure*}

\subsection{Effectiveness of Selection Algorithms}
We ran four selection algorithms. Namely, DPS, Greedy, Kascade, and GGA, to evaluate their effectiveness. We ran the GGA algorithm with only one pass and set both fission and fusion probabilities to one for a complete view set transformation. We tested the algorithms by varying the space budgets with S/6, S/4, and S/2, where S denotes the total view size $\sum_{v} s(v)$ of G-View in Table \ref{tab:view_overhead}. 


Figure \ref{fig:view_benefit} depicts the performance of selection algorithms in optimizing the view benefit. Overall, for any workload and space budget, the GGA algorithm achieved the highest view benefit, thus can have the largest query processing cost reductions. For the LDBC dataset with S/2, it improved 36\%, 20\%, 19\% of view benefit over DPS, Greedy, and Kascade, respectively. For the Amazon dataset with S/2, it achieved 30\%, 20\%, 9\% of view benefit improvement, respectively. The view benefit was significantly improved by GGA algorithm by 70\%, 150\%, 53\% in the DBLP dataset. Kascade had a higher benefit than DPS and Greedy because (i) it has eliminated the contained views, thus it selected more useful views than DPS, and (ii) it used the branch-and-bound strategy to search the solution, thus can optimize both view benefit and space. Nevertheless, it has an averagely 27\% lower benefit than GGA. This is mainly attributed to (1) GGA’s fine-grained view transformations that explore and merge the views with common subgraph parts. (2) its benefit evaluation strategy that can take multiple view combinations to optimize the benefit.

Figure \ref{fig:query_fraction} illustrates the fraction of queries that can be answered by the selected views. GGA clearly outperformed others because it employed supergraph views, merged views, and view combinations, which result in more contained queries. Particularly, it can fully cover all the queries when the space budget is increased to S/2. DPS had the lowest query coverage in the LDBC and Amazon datasets because it selected the views independently. Greedy had a higher query coverage than DPS because it removed the contained queries in each round. However, the query fraction of Greedy was affected by the low-utility views that have a high benefit and a large size in the DBLP dataset. Kascade can address this issue with its branch-and-bound solver, but still, it can not compete with GGA because it only considers single-view query rewriting. 


Figure \ref{fig:view_size} illustrates the size of candidate views generated by the selection algorithms. DPS had the largest view size because it had a candidate view for each query, while Greedy and Kascade had the same and relatively smaller size because they pruned the contained views. It is clearly visible that GGA method outperformed others because of its gene-based view transformation and combination. Particularly, it reduced the space of the view size by up to 61\%, 60\%, 58\% for LDBC, Amazon, and DBLP, respectively.



\subsection{Efficiency of Selection Algorithms}
\noindent Figure \ref{fig:running_time} shows the running time of four algorithms in milliseconds. In particular, the time consists of the execution time for view construction, view evaluation, and view selection. The results manifested that GGA outperformed others regarding efficiency. Overall, it accelerated 36\%, 20\%, 30\% of running time of DPS, Greedy, and Kascade for the LDBC workloads, respectively. The improvement was achieved up to 58\%, 71\%, and 55\% for the Amazon workloads, and 19\%, 59\% and 16\% for the DBLP workloads. Kascade was faster than DPS because it had a reduced candidate set after the view enumeration. Greedy incurred significant overhead because it had to re-evaluate the view benefit in each round. The primary advantage of GGA over others is that it has reduced the number and size of views in the candidate set, thus saved unnecessary computation of view evaluation. For the view selection phase, GGA was the best because it had the smallest candidate set to select and generate.



\subsection{The Convergence of GGA}
In this experiment, we investigated the convergence of GGA. We set both fission and fusion probabilities to 50\% and ran the algorithm with space budget S/2. The result was shown in Figure \ref{fig:EX3}c, which confirmed that GGA is effective: the algorithm converges within 10 generations for the workloads in three datasets. Furthermore, the results indicated that the strategy of state search is effective. When a state of view selection has a lower benefit than the previous state, the algorithm can jump to another state with a higher benefit.

\section{Related work}



\textbf{View selection for relational, XML and RDF data.} Materialized view selection in relational databases has been a well-studied topic (see \cite{ chirkova2012materialized, mami2012survey} for surveys). Particularly, Chaves et al. \cite{chaves2009towards} encoded the relational views as genes and applied the gene algorithm to the view selection problem in the setting of distributed databases. Recently, there emerged work, e.g., \cite{yuan2020automatic}, that utilized deep reinforcement learning to guide the view selection. There has been a host of work on processing XML queries using views \cite{katsifodimos2012materialized, mandhani2005query,  tang2009materialized}. In \cite{tang2009materialized}, the authors studied the view selection problem for XPath workloads, they proposed a greedy-based solution that makes the space/time trade-off. Katsifodimos et al. \cite{katsifodimos2012materialized} studied the view selection for XQuery workloads. They first developed a greedy-based algorithm for a Knapsack selection problem, then proposed a heuristic algorithm to search for an optimal view set based on multi-view rewriting. There has also been work for RDF view selection \cite{goasdoue2011view, castillo2010selecting}. Goasdou{\'{e}} et al. \cite{goasdoue2011view} solved the view selection problem as a search process. They proposed heuristic strategies to search for a set of reformulated RDF views to minimize the defined cost model. Unfortunately, none of these works considered the structural properties of graph queries in view selection, thus they cannot be applied directly to the graph view selection problem.

\textbf{View-based approaches in graph databases.} With the advances of graph databases, graph view-based approaches \cite{DBLP:conf/edbt/HassanKJAS18, tian2019synergistic, fan2014answering, da2019kaskade} have gained more and more attention. For instance, DB2 graph \cite{tian2019synergistic} utilized a graph overlay approach to define a graph view of the underlying relational data. Fan et al. \cite{fan2014answering} implemented graph views for pattern queries based on graph simulation \cite{fan2010graph}. GRFusion \cite{DBLP:conf/edbt/HassanKJAS18} decomposed the graph topology from the relational tables and used pointers to connect the graph topology with the relational attribute data. While the aforementioned methods implemented the graph view using the relational approaches, G-View proposed an extended graph view, which not only utilizes a native graph approach, but also supports the subgraph and supergraph query answering. Regarding view selection in graph databases, Fan et al. \cite{fan2014answering} studied the minimal and minimum containment problems but they considered the views were pre-computed and static, leading to duplicate view content. Kascade \cite{da2019kaskade} considered the view selection problem as an 0-1 Knapsack problem, which generated the candidates using constraint-based view enumeration, then used a branch-and-bound solver to select the views. Our work modeled the selection problem as an 0-1 Knasack problem as well. While Kascade only supported single-view rewriting, our GGA algorithm considered the subgraph/supergraph views, view transformations, and multi-view combinations, yielding a view set with a smaller view size and a higher view benefit.

\section{Conclusion}
In this work, we proposed an end-to-end tool, G-View, to automate the process of view selection in the graph databases. We proposed an \textit{extended graph view}, which can answer both the subgraph and supergraph queries. We devised a filtering-and-verification framework to check the query containment by views. We developed a search-based algorithm, GGA, which explores graph view transformations to reduce the view size and optimize the overall query performance. The experimental results manifested that G-View was significantly faster than other view-based methods in accelerating the queries while incurring smaller view overhead. Moreover, GGA outperformed other selection methods concerning effectiveness and efficiency. In the future, we plan to extend our techniques to other graph query languages such as Cypher \cite{Cypher} and SPARQL \cite{sparql11}.

{
	\bibliographystyle{abbrv}
	\bibliography{main}

\begin{thebibliography}{10}

\bibitem{Neo4j}
http://neo4j.com.

\bibitem{JanusGraph}
https://janusgraph.org/.

\bibitem{Cypher}
{Cypher: the Neo4j graph query Language}.
\newblock https://neo4j.com/cypher-graph-query-language/.

\bibitem{agrawal2000automated}
S.~Agrawal, S.~Chaudhuri, and V.~R. Narasayya.
\newblock Automated selection of materialized views and indexes in sql
  databases.
\newblock In {\em VLDB}, volume 2000, pages 496--505, 2000.

\bibitem{angles2017foundations}
R.~Angles, M.~Arenas, P.~Barcel{\'o}, A.~Hogan, J.~Reutter, and D.~Vrgo{\v{c}}.
\newblock Foundations of modern query languages for graph databases.
\newblock {\em ACM Computing Surveys (CSUR)}, 50(5):1--40, 2017.

\bibitem{beasley1993overview}
D.~Beasley, D.~R. Bull, and R.~R. Martin.
\newblock An overview of genetic algorithms: Part 1, fundamentals.
\newblock {\em University computing}, 15(2):56--69, 1993.

\bibitem{brocheler2011budget}
M.~Br{\"o}cheler, A.~Pugliese, and V.~S. Subrahmanian.
\newblock A budget-based algorithm for efficient subgraph matching on huge
  networks.
\newblock In {\em ICDE workshops}, pages 94--99. IEEE, 2011.

\bibitem{castillo2010selecting}
R.~Castillo and U.~Leser.
\newblock Selecting materialized views for rdf data.
\newblock In {\em International Conference on Web Engineering}, pages 126--137.
  Springer, 2010.

\bibitem{chaves2009towards}
L.~W.~F. Chaves, E.~Buchmann, F.~Hueske, and K.~B{\"o}hm.
\newblock Towards materialized view selection for distributed databases.
\newblock In {\em EDBT}, pages 1088--1099, 2009.

\bibitem{chirkova2003materializing}
R.~Chirkova and C.~Li.
\newblock Materializing views with minimal size to answer queries.
\newblock In {\em PODS}, pages 38--48. ACM, 2003.

\bibitem{chirkova2012materialized}
R.~Chirkova, J.~Yang, et~al.
\newblock Materialized views.
\newblock {\em Foundations and Trends{\textregistered} in Databases},
  4(4):295--405, 2012.

\bibitem{cordella2004sub}
L.~P. Cordella, P.~Foggia, C.~Sansone, and M.~Vento.
\newblock A (sub) graph isomorphism algorithm for matching large graphs.
\newblock {\em TPAMI}, 26(10):1367--1372, 2004.

\bibitem{da2019kaskade}
J.~M. da~Trindade, K.~Karanasos, C.~Curino, S.~Madden, and J.~Shun.
\newblock Kaskade: Graph views for efficient graph analytics.
\newblock In {\em ICDE}, 2020.

\bibitem{erling2015ldbc}
O.~Erling, A.~Averbuch, J.~Larriba-Pey, H.~Chafi, A.~Gubichev, A.~Prat, M.-D.
  Pham, and P.~Boncz.
\newblock {The LDBC social network benchmark: Interactive workload}.
\newblock In {\em SIGMOD}, pages 619--630. ACM, 2015.

\bibitem{fan2010graph}
W.~Fan, J.~Li, S.~Ma, N.~Tang, Y.~Wu, and Y.~Wu.
\newblock Graph pattern matching: from intractable to polynomial time.
\newblock {\em PVLDB}, 3(1-2):264--275, 2010.

\bibitem{fan2014answering}
W.~Fan, X.~Wang, and Y.~Wu.
\newblock Answering graph pattern queries using views.
\newblock In {\em ICDE}, pages 184--195. IEEE, 2014.

\bibitem{fletcher2017declarative}
G.~H. Fletcher, H.~Voigt, and N.~Yakovets.
\newblock Declarative graph querying in practice and theory.
\newblock In {\em EDBT}, pages 598--601, 2017.

\bibitem{goasdoue2011view}
F.~Goasdou{\'{e}}, K.~Karanasos, J.~Leblay, and I.~Manolescu.
\newblock View selection in semantic web databases.
\newblock {\em PVLDB}, 5(2):97--108, 2011.

\bibitem{gubichev2015query}
A.~Gubichev.
\newblock {\em Query Processing and Optimization in Graph Databases}.
\newblock PhD thesis, Technische Universit{\"a}t M{\"u}nchen, 2015.

\bibitem{gupta2005selection}
H.~Gupta and I.~S. Mumick.
\newblock Selection of views to materialize in a data warehouse.
\newblock {\em IEEE Transactions on Knowledge and Data Engineering},
  17(1):24--43, 2005.

\bibitem{DBLP:conf/edbt/HassanKJAS18}
M.~S. Hassan, T.~Kuznetsova, H.~C. Jeong, W.~G. Aref, and M.~Sadoghi.
\newblock Extending in-memory relational database engines with native graph
  support.
\newblock In {\em {EDBT}}, pages 25--36, 2018.

\bibitem{katsifodimos2012materialized}
A.~Katsifodimos, I.~Manolescu, and V.~Vassalos.
\newblock Materialized view selection for xquery workloads.
\newblock In {\em SIGMOD}, pages 565--576, 2012.

\bibitem{LdbcTechSpecification}
{LDBC task force}.
\newblock {The LDBC social network benchmark (version 0.3.2)}.
\newblock Technical report, Linked Data Benchmark Council, 2019.

\bibitem{lee2012depth}
J.~Lee, W.-S. Han, R.~Kasperovics, and J.-H. Lee.
\newblock An in-depth comparison of subgraph isomorphism algorithms in graph
  databases.
\newblock {\em Proceedings of the VLDB Endowment}, 6(2):133--144, 2012.

\bibitem{leskovec2007dynamics}
J.~Leskovec, L.~A. Adamic, and B.~A. Huberman.
\newblock The dynamics of viral marketing.
\newblock {\em TWEB}, 1(1):5--es, 2007.

\bibitem{leskovec2006patterns}
J.~Leskovec, A.~Singh, and J.~Kleinberg.
\newblock Patterns of influence in a recommendation network.
\newblock In {\em PAKDD}, pages 380--389. Springer, 2006.

\bibitem{lissandrini2018beyond}
M.~Lissandrini, M.~Brugnara, and Y.~Velegrakis.
\newblock {Beyond macrobenchmarks: microbenchmark-based graph database
  evaluation}.
\newblock {\em PVLDB}, 12(4):390--403, 2018.

\bibitem{mami2012survey}
I.~Mami and Z.~Bellahsene.
\newblock A survey of view selection methods.
\newblock {\em Acm Sigmod Record}, 41(1):20--29, 2012.

\bibitem{mandhani2005query}
B.~Mandhani and D.~Suciu.
\newblock Query caching and view selection for xml databases.
\newblock In {\em VLDB}, pages 469--480. VLDB Endowment, 2005.

\bibitem{SQLG}
P.~Martin.
\newblock {SQLG: an implementation of Apache TinkerPop on a RDBMS}.
\newblock http://sqlg.org/docs/2.0.0-SNAPSHOT/, 2020.

\bibitem{minot2015comparison}
M.~Minot, S.~N. Ndiaye, and C.~Solnon.
\newblock A comparison of decomposition methods for the maximum common subgraph
  problem.
\newblock In {\em 2015 IEEE 27th International Conference on Tools with
  Artificial Intelligence (ICTAI)}, pages 461--468. IEEE, 2015.

\bibitem{rodriguez2015Gremlin}
M.~A. Rodriguez.
\newblock The gremlin graph traversal machine and language (invited talk).
\newblock In {\em Proceedings of the 15th Symposium on Database Programming
  Languages}, pages 1--10, 2015.

\bibitem{sahu2017ubiquity}
S.~Sahu, A.~Mhedhbi, S.~Salihoglu, J.~Lin, and M.~T. {\"O}zsu.
\newblock The ubiquity of large graphs and surprising challenges of graph
  processing.
\newblock {\em PVLDB}, 11(4):420--431, 2017.

\bibitem{tang2008arnetminer}
J.~Tang, J.~Zhang, L.~Yao, J.~Li, L.~Zhang, and Z.~Su.
\newblock Arnetminer: extraction and mining of academic social networks.
\newblock In {\em SIGKDD}, pages 990--998, 2008.

\bibitem{tang2009materialized}
N.~Tang, J.~X. Yu, H.~Tang, M.~T. {\"O}zsu, and P.~Boncz.
\newblock Materialized view selection in xml databases.
\newblock In {\em DASFAA}, pages 616--630, 2009.

\bibitem{tarjan1972depth}
R.~Tarjan.
\newblock Depth-first search and linear graph algorithms.
\newblock {\em SIAM journal on computing}, 1(2):146--160, 1972.

\bibitem{tian2019synergistic}
Y.~Tian, W.~Sun, S.~J. Tong, E.~L. Xu, M.~H. Pirahesh, and W.~Zhao.
\newblock {Synergistic graph and SQL analytics inside IBM Db2}.
\newblock {\em PVLDB}, 12(12):1782--1785, 2019.

\bibitem{TinkerGraphdoc}
A.~Tinkerpop.
\newblock https://github.com/tinkerpop/blueprints/wiki/TinkerGraph.

\bibitem{Tinkerpopdoc}
A.~Tinkerpop.
\newblock https://tinkerpop.apache.org/docs/3.4.4/, 2020.

\bibitem{sparql11}
W3C.
\newblock {SPARQL 1.1 Overview}, 2013.

\bibitem{xquery}
W3C.
\newblock {XQuery 1.0: An XML Query Language (Second Edition)}, 2015.

\bibitem{yuan2020automatic}
H.~Yuan, G.~Li, L.~Feng, J.~Sun, and Y.~Han.
\newblock {Automatic View Generation with Deep Learning and Reinforcement
  Learning}.
\newblock ICDE, 2020.

\end{thebibliography}
}
\end{document}